\documentclass[11pt]{article}
\pdfoutput=1
\usepackage{graphicx,amsmath,amssymb}

\usepackage{color}
\usepackage{xcolor}
\definecolor{darkblue}{cmyk}{0.9,0.9,0,0}
\usepackage[colorlinks=true,linkcolor=darkblue,citecolor=darkblue,urlcolor=darkblue]{hyperref}
\usepackage{cite}
\usepackage{tcolorbox}
\usepackage{cases}

\graphicspath{{./plots/}{./}}

\usepackage{customcommands}
\usepackage{array}
\usepackage{enumerate}

\usepackage{empheq}

\begin{document}

\thispagestyle{empty}

\renewcommand{\thefootnote}{\fnsymbol{footnote}}
\setcounter{page}{1}
\setcounter{footnote}{0}
\setcounter{figure}{0}

\begin{flushright}
CERN-TH-2020-218
\end{flushright}

\begin{center}
$$$$

{\Large\textbf{\mathversion{bold}
Scattering from production in 2d
}\par}
\vspace{1.0cm}

\vspace{1.0cm}

\textrm{Piotr Tourkine$^{a,b}$ and Alexander Zhiboedov$^b$}
\\ \vspace{1.2cm}
\footnotesize{\textit{ 
$^a$CNRS, LPTHE, Sorbonne universit\'e, 4 place Jussieu, 75005 Paris, France\\
$^b$CERN, Theoretical Physics Department, CH-1211 Geneva 23, Switzerland\\
}
\vspace{4mm}
}

\par\vspace{1.5cm}

\textbf{Abstract}\vspace{2mm}
\end{center}

\noindent In 1968, Atkinson proved the existence of functions that satisfy all S-matrix axioms in four spacetime dimensions. His proof is constructive and to our knowledge it is the only result of this type. Remarkably, the methods to construct such functions used in the proof were never implemented in practice.
In the present paper, we test the applicability of those methods in the simpler setting of two-dimensional S-matrices.
We successfully implement two numerical iterative schemes (fixed-point iteration and Newton's method), which, by iterating unitarity and dispersion relations, converge to solutions to the S-matrix axioms. We characterize the region in the amplitude-space in which our algorithms converge, and discover a fractal structure connected to the so-called CDD ambiguities which we call ``CDD fractal''. To our surprise, the question of convergence naturally connects to the recent study of the coupling maximization in the two-dimensional S-matrix bootstrap. 
The methods exposed here pave the way for applications to higher dimensions, and expose some of the potential challenges that will have to be overcome.

\noindent

\vspace{1cm}

\hfill {\it Dedicated to the memory of Andr\'{e} Martin}

\numberwithin{equation}{section}

\setcounter{page}{1}
\renewcommand{\thefootnote}{\arabic{footnote}}
\setcounter{footnote}{0}

\setcounter{tocdepth}{2}

 \def\nref#1{{(\ref{#1})}}

\newpage

\tableofcontents

\parskip 1pt plus 1pt   \jot = 1.5ex

\newpage

\section{Introduction}
\label{sec:introduction}

Scattering and particle production are deeply connected in relativistic theories.\footnote{{By ``scattering'' we mean elastic scattering amplitudes $n \to n$. By ``production'' we mean inelastic scattering amplitudes $n \to m$.}} In two-dimensional spacetime, $d=2$, scattering without production leads to integrability \cite{Dorey:1996gd}. 
In higher dimensions, $d>2$, scattering without production is impossible \cite{aks1965proof}. In this paper we will be interested in the converse question: 
\begin{center}
{\it Given particle production, can we reconstruct scattering?}
\end{center}
In the simplest case of $2\to 2$ scattering, the familiar relation between the two is expressed by unitarity\footnote{For illustration purposes, we suppress the phase space integrals and simple kinematical pre-factors.}
\begin{equation}
  \label{eq:atk-schem}
\underbrace{ {\rm Disc}_s T_{2\to2}  -  | T_{2\to2} |^2 }_{\text{Scattering}} = \underbrace{ (\text{MP})  }_\text{Production}
\end{equation}
where $T_{2 \to 2}$ is the two-to-two scattering amplitude, ${\rm Disc}_  s$ stands for discontinuity with respect to Mandelstam variable $s$, and $({\rm MP})=\sum_{n>2} |T_{2\to n}|^2$ is the {\it multi-particle} contribution that describes particle production in a given scattering process. The nontrivial problem we are interested in here is to solve unitarity (\ref{eq:atk-schem}) combined together with analyticity and crossing (see section \ref{sec:definitions-review} for the precise formulation). 

Following the pioneering works by Atkinson \cite{Arkinson:1968hza,Atkinson:1969wy,Atkinson:1969eh,Atkinson:1970pe,Atkinson:1970zz}, we treat the particle production $({\rm MP})$ as a fixed, given data, and solve for $T_{2\to2}$ as a function of $({\rm MP})$. This is the central idea of the present work. In practice, we do this by implementing numerically an iterative procedure developed by Atkinson, which converges to a solution of \eqref{eq:atk-schem}.
Remarkably, given $({\rm MP})$ is not too large,\footnote{For $d=2$ the precise statement can be found in the bulk of the paper. For $d=4$ 
Atkinson proved sufficient conditions for convergence in \cite{Arkinson:1968hza,Atkinson:1969wy,Atkinson:1969eh,Atkinson:1970pe}.} the solutions to \eqref{eq:atk-schem} can be shown to exist and be reached by simple iteration procedures that we describe in detail below.

In this paper, we present and implement numerically two algorithms to reconstruct scattering from production in two spacetime dimensions. Our first method relies on applying iteratively the unitarity equation~\eqref{eq:atk-schem} to find a fixed-point thereof, our second method is the standard Newton method applied to this equation.

Two dimensions is an obvious starting point because S-matrices are much simpler: they have only one kinematic invariant, and no phase-space integral in the unitarity equation. Furthermore, for scattering of identical particles the problem can be solved analytically and therefore the performance and limitations of the numerical algorithms can be explored in great depth. While in two dimensions our algorithm simply recovers known solutions in a novel way, its power lies in the fact that it can be easily generalized to higher dimensions where no solutions are known.

We call a function of Mandelstam invariants that satisfies the fundamental constraints  of analyticity, unitarity, and crossing an amplitude-function. It is one of the aims of the nonperturbative S-matrix program
to characterize the space of such functions.\footnote{Note that the basic principles discussed here are known to lead to many nontrivial constraints on the physical parameters of the theory, see \cite{Bellazzini:2020cot,Tolley:2020gtv,Caron-Huot:2020cmc,Arkani-Hamed:2020blm} for some recent works. In these works, however, only some of the constraints coming from unitarity are implemented. In the problem we are considering we aim at imposing unitarity fully at the level of two-to-two scattering amplitude in a gapped theory (no massless particles in the spectrum).} It does not necessarily describe scattering in some physical theory, but the converse is definitely true: every physical scattering amplitude is an amplitude-function. Constructing amplitude-functions and characterizing the space of physical parameters that they span, which we can loosely call the space of couplings, is an interesting and important problem, sometimes also called the {\it primal} problem, see \cite{Guerrieri:2020kcs} for the recent discussion of both the primal and dual problems in 2d. Even more ambitiously, by exploring the space of the amplitude-functions, we might hope to find physical theories at its boundary and in this way try to solve them.

An important step in this direction has been taken in \cite{Paulos:2016fap,Paulos:2016but,Paulos:2017fhb,Homrich:2019cbt,Guerrieri:2020bto}, where inelastic unitarity constraints were fully implemented and nonperturbative constraints on the scattering amplitudes derived. Remarkably, in two spacetime dimensions this led to re-discovering the actual physical S-matrices using purely bootstrap methods. Finding such exact, solvable S-matrices in higher dimensions is still elusive.

To complete the program of constructing amplitude-functions in higher dimensions, one extra condition has still to be implemented for massive theories: \textit{elastic unitarity}. Elastic unitarity is the statement of unitarity at energies where no particle production is possible. In this energy range (e.g. $4m^2<s<16m^2$ for pion scattering), ${\rm (MP)}$ vanishes in \eqref{eq:atk-schem}. When combined with analyticity and crossing, elastic unitarity is known to give many powerful and nontrivial constraints on the scattering amplitudes, see e.g. \cite{Correia:2020xtr} for a recent discussion of the question.

The methods that have been developed so far do not allow to impose elastic unitarity, due to its nonlinear nature. It is therefore an open problem to explore the space of amplitudes that satisfy both elastic and inelastic unitarity, analyticity and crossing.

In this context, one remarkable result that so far has attracted little attention is a constructive proof by Atkinson \cite{Arkinson:1968hza,Atkinson:1969wy,Atkinson:1969eh,Atkinson:1970pe} of the \textit{existence} of functions satisfying all of the S-matrix axioms: crossing, elastic and inelastic unitarity, in four dimensions. 
The mathematical proof is based on constructing an iterative sequence and exhibiting some sufficient conditions about its convergence. To our knowledge (see also \cite{Kupsch:2008hq}), it is the only result about two-to-two scattering amplitudes in gapped theories that has all the desired unitarity properties.

The main message of our paper is that the approach of imposing unitarity advocated in the papers by Atkinson is actually suitable for efficient numerical implementation. It therefore offers an exciting avenue to explore the space of scattering amplitudes that satisfy all the desired unitarity properties in higher dimensions and overcome some of the present limitations of the S-matrix program.  If implemented in four dimensions, it would be the first method able to construct amplitude-functions which satisfy all axioms of the S-matrix program and therefore we believe that this idea deserves further attention \cite{WIP}.

From a more philosophical standpoint, \textit{choosing} $({\rm MP})$ seems to imply a huge amount of indeterminacy on the resulting scattering amplitude-function. Nevertheless, we see two main aspects that justify this approach. Firstly, when deriving bounds on the space of low-energy S-matrix parameters, we expect that detailed form of the multi-particle input to be irrelevant. By elevating the $({\rm MP})$ particle production amplitude to the role of an input parameter and observing how the $2\to2$ S-matrix depends on it by varying its parameters, we can very explicitly test this idea. Secondly, to bootstrap particular physical theories, we have the possibility to input particle production data obtained by other means and in this way narrow the search for the desired amplitude in the potentially vast space of amplitude-functions. Thirdly, it provides an interesting middle ground between perturbative and nonperturbative methods. In solving \eqref{eq:atk-schem}, we can use for example perturbative methods to get some insight into the precise form of the particle production term $({\rm MP})$, which can be then turned into the scattering amplitude using the nonperturbative methods of the present paper.

The main results of this paper are
\begin{itemize}
\item We implemented numerically for the first time 
iterative solutions to unitarity, analyticity, and crossing with a given inelasticity $({\rm MP})$ in two space-time dimensions. We recovered the known integrable S-matrices and analogues with inelasticity which match the known solutions to unitarity in 2d.
\item We characterized in details the range of convergence of the two methods we used to solve (\ref{eq:atk-schem}): fixed-point iteration, described in section \ref{sec:atkinson-numerics}; Newton's method, described in section \ref{sec:newt-kant-meth}. We see that most of the parameter space of theories can be captured, appart from a slice near the edge, for which our algorithms do not converge. Covering the complete space of solutions in two space-time dimensions requires methods that go beyond the ones described in the paper.

\item In $d=2$, particle production $({\rm MP})$ does not specify the scattering amplitude $T_{2 \to 2}$ uniquely. In fact, there are infinitely many scattering amplitudes $T_{2 \to 2}$ with a given inelasticity. This space of solutions to our problem are characterized by  CDD ambiguities \cite{Symanzik:1961,Creutz:1973rw,Mussardo:1999aj}. We explored how different starting points of Newton's method allow to recover S-matrices with different CDD factors.
\end{itemize}

The plan of the paper is as follows. In section~\ref{sec:atkinson-maps}, we review the basics of S-matrices in 2d and the analytic solutions which we will recover through our numerical implementation. In section~\ref{sec:atkinson-numerics} we describe our numerical implementation of the fixed-point iteration in two dimensions, and in section~\ref{sec:newt-kant-meth} we explain how we used Newton's method to answer the question posed at the beginning of the paper.
Section~\ref{sec:discussion} presents the discussion of the results and some open directions.

\section{Two-dimensional S-matrices}
\label{sec:atkinson-maps}

In this section, we review the basic properties of two-dimensional S-matrices and formulate precisely the problem of scattering from production that we would like to solve. We review the analytic solution to the problem which involves the notion of CDD S-matrices \cite{Mussardo:1999aj}, central to this work. Then, we present our numerical iterative solution to the problem which is the subject of the present paper.

\subsection{Problem}
\label{sec:definitions-review}

We consider scattering of identical massive bosons in two spacetime dimensions. The two-to-two kinematics is captured by one kinematic invariant, the center of mass energy $s$, while $t=0$ and $u=4m^2-s$. The $S$-matrix is therefore a function of $s$ only the scattering amplitude $T$ is defined by
\be
S(s) = 1 + i {T(s) \over \sqrt{s(s-4 m^2)}} .
\ee
In the free theory, $T(s)=0$. The inverse factor of $ \sqrt{s(s-4 m^2)}$ comes from the Jacobian that relates $\delta^{(2)}(p_1+p_2+p_3+p_4)$ to $\delta^{(2)}(p_1+p_3) \delta^{(2)}(p_2+p_4)$ in two dimensions (see for instance  \cite{Vieira:TASI}).

We assume that $S(s) \equiv \lim_{\epsilon \to 0} S(s + i \epsilon)$ satisfies Mandelstam analyticity, namely that it is holomorphic outside of the unitarity cuts $s<0$ and $s \geq 4 m^2$, apart from potential poles that correspond to bound states located at $0 < s< 4 m^2$. Real analyticity implies that $T(s^*)=T^*(s)$. Crossing symmetry takes the form
\be
\label{eq:crossing}
S(s) = S(4 m^2 - s) .
\ee
Finally, assuming that particle production starts from $s_0$, unitarity takes the following form
\be
\label{eq:elasticunitarity}
|S(s)| &=1 , ~~~ 4 m^2 \leq s < s_0 \ ,  \\
|S(s)| &\leq 1 , ~~~ s \geq s_0 \ .
\label{eq:unitarity}
\ee
In a theory with a single stable particle of mass $m$, $s_0 = 9 m^2$ and $s=16 m^2$ stands for the three- and four-particle thresholds correspondingly. The explicit value of $s_0$ does not play an important role in the subsequent analysis.

To characterize particle production, we explicitly introduce inelasticity as follows
\be
S(s) S^*(s) 
\equiv 1 - f_{i}(s) \geq 0 ,
\label{eq:finel-def}
\ee
where $f_i(s)$ is defined for real, positive $s$ and by assumption it has a nontrivial support starting for $s \geq s_0$. It characterizes the amount of particle production. Below, for concreteness we set $s_0 = 16 m^2$ which corresponds to a situation where $T_{2 \to 3} = 0$.

In terms of $T(s)$, unitarity and inelasticity read
\begin{equation}
  \label{eq:T-unit}
\Im T(s) = \frac1{2\sqrt{s(s-4m^2)}}|T(s)|^2 + v_i(s) \ ,
\end{equation}
where
\begin{equation}
  \label{eq:vi-fi}
 v_{i} (s) =  f_i(s)   \frac{ \sqrt{s (s-4m^2)}}{4} . 
\end{equation}

At this point we can formulate precisely the problem that we would like to solve.

\begin{tcolorbox}
 {\bf Problem:} Given inelasticity $v_i(s)$, find the scattering amplitude $T(s)$ that satisfies Mandelstam analyticity, crossing \eqref{eq:crossing},  elastic unitarity \eqref{eq:elasticunitarity}, and inelastic unitarity \eqref{eq:unitarity}. 
 \end{tcolorbox}

\subsection{Explicit solution}
\label{sec:solution}
It turns out that an explicit solution to the problem stated above exists in two dimensions. We consider the cases $v_i(s) =0$ and $v_i(s) \neq 0$ separately.

\paragraph{Elastic amplitudes ($v_i(s) = 0$)}

In two dimensions, it is possible to have scattering without particle production. Such theories are integrable, and they are characterized by an S-matrix which is a pure phase. The building blocks thereof are known as CDD factors. Following the terminology of \cite{Paulos:2016but}, we will distinguish CDD poles and CDD zeros.

The corresponding S-matrices are given by
\begin{align}
S_{\text{CDD}}^{\text{pole}}(s) &= { \sqrt{s (s-4 m^2)} + \sqrt{m_p^2 (4 m^2 - m_p^2)} \over \sqrt{s (s-4 m^2)} - \sqrt{m_p^2 (4 m^2 - m_p^2)}}\,,\\ 
S_{\text{CDD}}^{\text{zero}}(s) &= { \sqrt{s (s-4 m^2)} - \sqrt{m_{z}^2 (4 m^2 - m_{z}^2)} \over \sqrt{s (s-4 m^2)} + \sqrt{m_{z}^2 (4 m^2 - m_{z}^2)}} .
\end{align}
These S-matrices satisfy Mandelstam analyticity, crossing, and unitarity $| S_{\text{CDD}}^{\text{pole}}(s)| = | S_{\text{CDD}}^{\text{zero}}(s)  | = 1$. It is easy to check that $S_{\text{CDD}}^{\text{pole}}(s)$ has a pole at $s=m_p^2$, and that $S_{\text{CDD}}^{\text{zero}}(m_z^2) = 0$. 

It is also possible for $S(s)$ to have a pair of complex conjugate zeros.\footnote{In ~\cite{Paulos:2016but} such pairs of zeros were called CDD resonances.}  This can be achieved by either choosing a single CDD factor with $m_z^2 = 2 m^2 + i \alpha$ with $\alpha\in \mathbb{R}$, or by taking a product of two $S_{\text{CDD}}^{\text{zero}}$ with complex conjugate zeros. 

It is easy to see that any purely elastic S-matrix in 2d is a product of CDD-poles and CDD-zeros. Firstly, map one half of complex plane ($\Re(s)>2$) minus the right cut $s>4m^2$ to the unit disk via $s\to z= { \sqrt{s (s-4 m^2)} + \sqrt{a (4m^2 - a)} \over \sqrt{s (s-4 m^2)} - \sqrt{a (4 m^2 - a)}}$ for some $a$ (see \cite[fig.~1]{Paulos:2017fhb}). Suppose that there exists an S-matrix $\tilde S(z)$ with poles and zeros at some location in the complex plane, and define $f(z)$ to be the ratio of $\tilde S(z)$ to the CDD-pole and CDD S-matrices with the same poles and zeros. $f(z)$ is therefore holomorphic, and since it possesses no zeros in the unit disk, $1/f(z)$ is also holomorphic. Since the cut maps the left cut $s>4m^2$ to the boundary of the unit disk (again, see \cite[fig.~1]{Paulos:2017fhb}), $|f(\exp(i\theta))|=1$ for $\theta\in[0;2 \pi]$. The application of the maximum modulus principle to $f$ and $1/f$ gives that $f(z)=1$ everywhere. Crossing symmetry allows to extend this to the full complex plane.

\paragraph{Inelastic amplitudes ($v_i(s) \neq 0$)}

Following \cite{Paulos:2016but,Creutz:1973rw}, we can immediately write down a general solution to eq.~\eqref{eq:T-unit}, as follows:
\begin{equation}
\label{eq:solution}
S(s)=S_{\text{elastic}}(s) e^{\int_{4m^2}^\infty {ds' \over 2 \pi i} \log(1 - f_{i}(s')) 
    \sqrt{s(s-4 m^2) \over s'(s'-4 m^2)} 
  \left( {1 \over s' - s} + {1 \over s' - (4 m^2 - s) }  \right)} ,
\end{equation}
where $|S_{\text{elastic}}(s)|=1$ is a purely elastic $S$-matrix given by the product of CDD zeros and CDD poles, as well as possibly an overall sign factor. It is immediate to check that (\ref{eq:solution}) satisfies all the conditions described previously. Note that this formula assumes that $|\log \left(1 - f_{i}(s) \right)| < s$ at large $s$. This will be enough for the purposes of this paper.\footnote{By changing the power in (\ref{eq:solution}) to $\left( {s(s-4 m^2) \over s'(s'-4 m^2)} \right)^{{1\over 2} + n}$, we can generalize this formula to describe $|\log(1 - f_{inel}(s))| < s^{1+2n}$. Presumably, this induces new elastic terms in the form of CDD zeros, which distinguish between different amplitudes with different $n$'s.
}

\paragraph{Properties of $T(s)$}

Below we will be interested in the properties of the scattering amplitude $T(s)$, in addition to those of $S(s)$. Let us list a few of its relevant properties which concern its behavior close to the two-particle threshold $s=4m^2$ as well as its high energy limit $s \to \infty$. 

Consider an amplitude with $N_{zeros}\geq0$ CDD zeros, $N_{poles}\geq0$ CDD poles and inelasticity (possibly zero). For $s \to \infty$, this amplitude goes to a constant
\begin{equation}
\lim_{s \to \infty} T(s) :=c_{\infty} \label{eq:constant-def}
\end{equation}
given by (the computation is straightforward):
\begin{align}
  \label{eq:constant}
c_{inf} &= 2\sum_{i=1}^{N_{poles}}\sqrt{m_{p_i}^2(4 m^2-m_{p_i}^2)} - 2\sum_{j=1}^{N_{zeros}}\sqrt{m_{z_j}^2(4 m^2-m_{z_j}^2)} \nn \\
                         &+ \int_{4m^2}^\infty {ds' \over \pi} \log[1 - f_{i}(s')] \frac{1}{\sqrt{ s'(s'-4 m^2)}} 
                           \left( s' - 2 m^2 \right) .
\end{align}
Close to the threshold $s\simeq 4m^2$, the behaviour of the amplitude depends on wheter the total number of CDD poles and zeros $N_{tot}= N_{zeros} + N_{poles}$ is even or odd:
\begin{equation}
  \label{eq:near-threshold}
\begin{aligned}
    T(s) &= 4 i m \sqrt{s-4m^2} + O(s-4) , ~~~ N_{tot} \text{~~~ even}, \\  
  T(s) &= c_0 i (s-4m^2)^{{3 \over 2}} + O((s-4)^2) , ~~~ N_{tot} \text{~~~ odd} .
\end{aligned}
\end{equation}
where $c_0$ is a real coefficient depending on the location of the zeros and poles whose explicit value is not important for us but straightforward to work out.

\subsection{Iterative solution}
\label{sec:atkinson-solver}

Next, we present an iterative solution to the problem above that we implement numerically in the following sections. Following the basic idea of Atkinson, we define the iterative map $\Phi$ as the right-hand side of the unitarity equation
\begin{equation}
  \label{eq:phi-def}
  \Phi(\Im T)(s) = \frac{1}{2\sqrt{s(s-4 m^2)}}|T(s)|^2+v_i(s) .
\end{equation}
The RHS (right-hand side) of \eqref{eq:phi-def} explicitly depends on the real part of the amplitude: using dispersion relations,
\begin{equation}
  \label{eq:equiv-disp}
  T(s) = c_\infty -\frac{g^2}{s-m_p^2}-
  \frac{g^2}{4 m^2 -s-m_p^2}
  + \int_{4 m^2}^\infty {d s' \over \pi} \Im T(s')
  \left(\frac{1}{s'-s}+\frac{1}{s'-(4 m^2-s)}\right)\,,
\end{equation}
it can be expressed in terms of the imaginary part of the amplitude $\Im T(s)$. This fact justifies writing $\Phi(\Im T)(s)$. Here, for simplicity we assumed that only a single bound state is present in the spectrum. Analogous formulas can be easily written for an arbitrary number of bound states.

The first iterative strategy that we consider is therefore to define a sequence
\begin{equation}
  \label{eq:Tn-FP}
  \Im T_{n+1} = \Phi(T_n) \ ,
\end{equation}
which, whenever it converges, converges to a fixed-point $\Im T=\Phi(\Im T)$ that, by construction, satisfies unitarity. Crossing is ensured by \eqref{eq:equiv-disp}.

Our second iterative strategy is the Newton-Kantorovich method, a functional analogue of Newton's method, which we apply to find the roots of the map 
\be
\Psi \equiv \mathrm{id}-\Phi \ ,
\label{eq:psi-def}
\ee
where $\mathrm{id}$ is the identity functional. In that case we look at the following iteration:
\begin{equation}
  \label{eq:NK}
    \Im T_{n+1}= \Im T_{n} - (\Psi')^{-1} \cdot \Psi(T_n) \ ,
\end{equation}
where $\Psi'$ is the functional derivative of $\Psi$ with respect to $\Im T$. The roots of this operator clearly also satisfy unitarity and crossing.
We make this more explicit in the section dedicated to Newton's method below. Note also that in our implementation, we deal with a discretized version of this algorithm, which is simply called ``Newton's method''.

As we mentioned above, the map $\Phi$ uses explicitly {involves} the real part of the amplitude. Therefore, it {contains an implicit step} of reconstructing the real part, from the imaginary part via dispersion relations which can be recast as
\begin{equation}
  \label{eq:Re-disp}
     \Re T_{n}(s) = c_\infty -\frac{g^2}{s-m_p^2}-
   \frac{g^2}{4 m^2 -s-m_p^2}
   + P.V. \int_{4 m^2}^\infty {d s' \over \pi} \Im T_{n}(s')
   \left(\frac{1}{s'-s}+\frac{1}{s'-(4 m^2-s)}\right) ,
\end{equation}
where taking the real part prescribes the integral to be understood in the sense of its principal value.

In a practical implementation, the challenge that one faces is that $T_{n}(s)$, and consequently $\Re T_n(s)$, should vanish when $s\to 4 m^2$ (see \eqref{eq:near-threshold}), but the dispersion relation above does not guarantee this vanishing. There are two strategies to enforce this condition, which we describe now.

Most of the time, in our algorithm, by convention, we will keep the constant at infinity $c_{\infty}$ fixed throughout the iteration and consider it an input of the algorithm. We will then use the vanishing of $\Re T_{n}(4m^2)=0$ to \textit{define} a sequence of couplings $g_n$ according to
\begin{equation}
  \label{eq:gn-continuous}
  g_n^2 = \left(\frac1{4m^2-m_p^2}-\frac1{m_p^2}\right)^{-1}\left(c_\infty+ \int_{4 m^2}^\infty {d s' \over \pi} \Im T_{n}(s')
   \left(\frac{1}{s'-4m^2}+\frac{1}{s'}\right) \right) .
\end{equation}
Note that this result is actually generic: for an amplitude with only one bound state, the coupling is given by the dispersion integral above.
Generalizing to more bound states is straightforward. One can for instance, following \cite{Paulos:2016but}, keep all fixed but one and define $g_n$ in a very similar maner, by adding the contribution from the other poles at $s=4m^2$ in the parenthesis on the right-hand side of \eqref{eq:gn-continuous}.

However, in the cases where there are no poles, we cannot keep $c_\infty$ fixed and we have to define a sequence of constants at infinity $c_n$ by
\begin{equation}
  \label{eq:cn-continuous}
  c_n =-  \int_{4 m^2}^\infty {d s' \over \pi} \Im T_{n}(s')
  \left(\frac{1}{s'-4m^2}+\frac{1}{s'}\right) .
\end{equation}

To sum up, in the case where we have one bound state, we construct the iteration procedure as follows:
\begin{tcolorbox}
{\bf Iterative solution:}
\begin{subequations}
\begin{empheq}[left={\Im T_{n+1}(s) = \empheqlbrace\,}]{alignat=2}
  \Phi(\Im T_{n}) \hspace{66pt}  & ~~~~~\text{(fixed-point iteration)} \label{eq:fixedpointiter} \\
\Im T_{n} - (\Psi')^{-1} \cdot \Psi(T_n) &~~~~~ \text{(Newton-Kantorovich method)} \label{eq:NKmethod}
\end{empheq}
\label{eq:schem-atk1}
\end{subequations}
\begin{align}
  \label{eq:schem-atk2}
    T_{n+1}(s) &= c_\infty -\frac{g_{n+1}^2}{s-m_p^2}-
   \frac{g_{n+1}^2}{4 m^2 -s-m_p^2}
   + \int_{4 m^2}^\infty {d s' \over \pi} \Im T_{n+1}(s')
   \left(\frac{1}{s'-s}+\frac{1}{s'-(4 m^2-s)}\right)  \\ 
     \label{eq:schem-atk3}
      g_{n+1}^2 &= \left(\frac1{4m^2-m_p^2}-\frac1{m_p^2}\right)^{-1}\left(c_\infty+ \int_{4 m^2}^\infty {d s' \over \pi} \Im T_{n+1}(s')
   \left(\frac{1}{s'-4m^2}+\frac{1}{s'}\right) \right) 
   \end{align}
\end{tcolorbox}

The input data required for the iteration procedure is:
\begin{itemize}
\item mass of the bound state $m_p$,
\item inelasticity $v_i(s)$, 
\item constant at infinity $c_\infty$.
\end{itemize}
Given the input data, and an arbitrarily chosen initialization amplitude $T_0(s)$ we can iterate the three equations above. Unitarity relation 
allows us to compute the discontinuity of the amplitude at the next step of the iteration. The essential step of this procedure is the reconstruction of the real part of $T_{n}(s)$ via dispersion relations \eqref{eq:schem-atk2}. The condition $T_n(4m^2)=0$ \eqref{eq:schem-atk3}, which is a consequence of elastic unitarity \eqref{eq:elasticunitarity}, guarantees that the dispersion integral is well-defined at every iteration. Depending on our choice of the problem we can alternatively fix $g^2$ and iterate $c_\infty$ instead.

In implementing the iterative solution above, we will have to address two principal questions:
\begin{itemize}
\item when does a given numerical strategy converge?
\item what does it converge to, as a function of the input data and the initialization amplitude $T_0(s)$?
\end{itemize}
These questions will be the subject of the following sections dedicated to the numerical implementation of the iterative algorithm.

\subsection{Ambiguities in the reconstruction}

Given the input data, and looking back at the explicit solution \eqref{eq:solution} it is immediate to see that the ineslastic contribution is fully determined, and so is the pole contribution of $S_\mathrm{elastic}$. However, it is not difficult to realize that a solution of interest can have an arbitrary number of CDD zeros, as long as they are arranged so that the sum of \eqref{eq:constant} is respected, i.e.
\be
\label{eq:locationzero-c-inf}
c_{\infty} &=2\sqrt{m_{p}^2 (4 m^2 -m_{p}^2)} - 2\sum_{j=1}^{N_{zeros}} \sqrt{m_{z_j}^2 (4 m^2 -m_{z_j}^2)} \nn \\
 &+ \int_{4m^2}^\infty {ds' \over \pi} \log[1 - f_{i}(s')] \left( {1 \over s'(s'-4 m^2)} \right)^{{1 \over 2}} \left( s' - 2 m^2 \right) .  
\ee

Whenever convergent, we will find below that the fixed-point iteration algorithm converges to the $N_{zeros}=1$ solution independently of the initialization amplitude $T_0(s)$. On the other hand, we will also see that the Newton method can converge to arbitrary $N_{zeros}$ depending on the initialization amplitude $T_0(s)$. Finally, there are amplitudes for which neither of the methods converge and some other method should be used.

A related source of ambiguity is related to the fact that, at finite numerical resolution,\footnote{Namely working on a given discretized grid with some fixed numerical precision.} one cannot distinguish amplitudes which differ by an insertion of even number of CDD zeros arbitrarily close to $s=4 m^2$. This ambiguity will not play any role in our discussion below, but the reader should keep in mind that adding such undetectable CDD zeros is always formally possible in going from the discretized problem back to the continuum.

\section{Fixed-point iteration}
\label{sec:atkinson-numerics}

In this section, we implement numerically the fixed-point iteration algorithm described above, see \eqref{eq:fixedpointiter}. Remarkably, we find a wide class of amplitudes for which the algorithm converges.\footnote{Stated differently we can say that we find that the iterative map \eqref{eq:fixedpointiter} is contracting. For the definition of contracting maps and review of the corresponding fixed point theorems, see e.g. \cite{Atkinson:1970zza}.} Below and for the rest of the paper, we will set $m=1$. It will also be convenient to map the domain $s\in[4;+\infty)$ to $(0,1]$ via
\begin{equation}
x=4/s\label{eq:x-def} \ .
\end{equation}
To avoid confusion, we should use a different name for functions of $x$, for instance $\tilde T(x) \equiv T(4/x)$. Hoping that it will not confuse the reader, below, we omit tildes and simply write $T(x)$. In this variable, the two-particle threshold at $s=4$ is mapped to $x=1$, infinite $s$ is mapped to $x=0$, and the four-particle inelastic threshold at $s=16$ maps to $x=1/4$. Finally, we introduce for the imaginary part of the amplitude
\be
\rho(x) \equiv \Im T(x) \,,\quad x\in [0,1]\,.
\ee

The steps of the numerical implementation are as follows:
\begin{enumerate}
\item {\it Discretization.} We discretize the values of the function that we iterate on a grid of length $N$
  \begin{equation}
x_0=0<x_1\cdots x_{N-1}<x_N=1\,.\label{eq:grid}
\end{equation}
The grid spacing controls the precision of the overall agreement with the analytic solutions. 
\item {\it Interpolation.} We adopt two methods of interpolation: linear interpolation, and interpolation with Bernstein polynomials, see respecively appendices \ref{sec:linear-interpolation} and \ref{app:bernstein}. Linear interpolation allows an easier experimentation with various grids and local increase of precision near points of interest. The grid we used for the results of this paper for linear interpolants is given in \eqref{eq:grid-def}. Bernstein polynomials require a unformly-spaced grid but result in smoother functions and are known to converge uniformly upon increasing the resolution of the grid. The value of the function $\rho_n$ at step $n$ of the iteration on the grid are called $\rho_{n,i}$
  \begin{equation}
    \label{eq:rhoi-def}
    \rho_n(x_i)=\rho_{n,i},\quad \rho_{n,0}=0,~\rho_{n,N}=0\,.
  \end{equation}
  Here $x_0=1$ and $x_N=1$.

\item {\it Dispersion integral \eqref{eq:schem-atk2}.} The dispersion integral reconstructs the real part of $T(x)$ from its imaginary part obtained using  \eqref{eq:schem-atk2}. Given a grid and an interpolation scheme, this dispersion integral reduces to following matrix operation on $\rho_n$, just like in \cite{Paulos:2016but},
\begin{equation}
  \label{eq:re-discrete}
\Re T_{n,i} = c_\infty -g^2_n\left(\frac1{4/x_i-m_p^2}-\frac1{4/x_i-(4-m_p^2)}\right)+\frac 1 \pi\sum_{j=1}^{N-1} B_{ij}\rho_{n,j} .
\end{equation}
The reconstruction-matrix $B_{i j}$ is computed once and for all, given the discretization and interpolation schemes. The explicit form of this matrix for the linear interpolation and interpolation with Bernstein polynomials can be found in the dedicated appendices.

The coupling $g_n$ in \eqref{eq:re-discrete} is defined by demanding that $\Re T_{n,N} =0$ (recall that $x_N=1$ corresponds to $s=4$):
\begin{equation}
  \label{eq:gn}
  g_n^2 = \left(\frac1{4-m_p^2}-\frac1{m_p^2}\right)^{-1}\left(\frac1 \pi\sum_j B_{Nj}\rho_{n,j}+c_\infty\right) .
\end{equation}
This is the discretized version of \eqref{eq:schem-atk3}.
Once this is done, $\Re T_{n}$ reads
\begin{equation}
  \label{eq:Re-form}
  \Re  T_{n,i}=G_{i,j} \rho_{n,j}+q_i
\end{equation}
where
  \begin{equation}
    \label{eq:G-def}
    G_{n,ij}=
      B_{ij}-\frac{P(x_i)}{P(1)} B_{Nj} 
    ,,\quad
  q_i = c_\infty\left(1-\frac{P(x_i)}{P(1)}\right)\,,
\end{equation}
and $P(x)$ is the pole function in the $x$-variable:
\begin{equation}
  \label{eq:pole-def}
  P(x) \equiv \frac1{4/x-m_p^2}-\frac1{4/x-(4-m_p^2)} .
\end{equation}

\item {\it Iteration.} The resulting map assumes the final form
  \begin{equation}
    \rho_{n+1,i} = \Phi(\rho_{n,j})_i:=\frac{x_i}{8\sqrt{1-x_i}} \Big(\rho_{n_i}^2+ ( \sum_j G_{ij}\rho_j+q_i )^2 \Big)+v_{i}(x_i)
      \label{eq:map-discrete}
  \end{equation}
  This is the discretized version of  \eqref{eq:fixedpointiter}--\eqref{eq:schem-atk3}.
\end{enumerate}

With the explicit form of the map given in \eqref{eq:map-discrete} at hand, we can apply and implement numerically theorems about the convergence of such maps. In finite dimensions, a map $\vec \rho{\,}' = \Phi(\vec \rho)$ converges locally to a fixed point $\vec \rho{\,}^* = \Phi(\vec \rho{\,}^*)$ if the largest eigenvalue of $\Phi$  \textit{at the fixed point}, also called \textit{spectral radius}, is strictly smaller than one. This result is standard and easily proven. Upon iteration, in a small neighborhood of the fixed point, any large eigvenvalue will drive the iteration away and small eigvenvalues will make it converge.\footnote{Of course if, for instance, the matrix has blocks which are not contributing, one can refine this statement.} 

In our case, the Jacobian of the map can be computed explicitly and is given by
\begin{equation}
  \label{eq:discrete-Jac}
  J_{n,ij} = \frac{\partial \rho_{n,i}}{\partial \rho_{n,j}} = \frac{x_i}{4\sqrt{1-x_i}} \Big( \rho_{n,i}\delta_{i,j} + G_{ij}(G_{ik}\rho_{n,k} + q_i) \Big), ~~~ i,j=1, .., N-1 ,
\end{equation}
where $i$ is not summed. The eigenvalues of the Jacobian are easily computed numerically, and we verify very accurately that, when parameters are such that the spectral radius of the iteration is bigger than one, the algorithm diverges, while when it converges it does so at the exponential rate of the largest eigenvalue.\footnote{This is so, because the quantities we look at, for instance the coupling or constant at infinity are not eigenvectors of this Jacobian, hence they receive contributions from all modes.}

We now present our results, and start with a few examples of iterations.

\subsection{Convergent amplitudes}
\label{sec:conv-one-bound}

Here we present examples of the amplitudes for which the iteration converges. In terms of the analytical solution \eqref{eq:solution} we consider mainly two cases: no bound state (and no CDD zeros), one bound state (and one CDD zero). We have also determined the convergence criterion of the algorithm in these cases, which we present below.

\subsubsection{No bound states}

We start with the case where we have no bound state below the two-particle threshold $s=4$. This is a small subset of the amplitudes we study, one for which $S_\mathrm{elastic}=1$ in \eqref{eq:solution}. For them, convergence is easy to understand, and the algorithm simply stops converging when $f_i>1$, because there are no such S-matrices, as evident from eq.~\eqref{eq:finel-def}.
Therefore, we need to have some inelasticity, otherwise the algorithm simply converges to zero. 

Besides, since we have no bound state, we need to turn temporarily to the other strategy described around \eqref{eq:cn-continuous}, and adjust the constant at infinity at each step of the iteration so as to cancel the real part at $s=4$. 

\begin{figure}[h]
  \hspace{-24pt}   
  \begin{tabular}{m{5cm}  m{5cm} m{5cm}}
\includegraphics{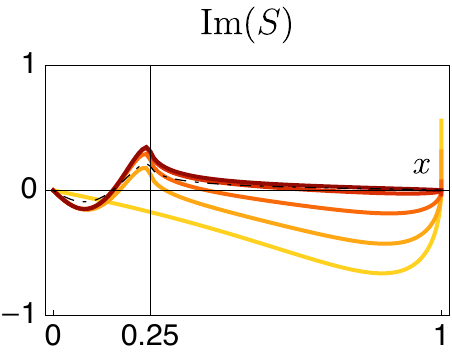} &
\includegraphics{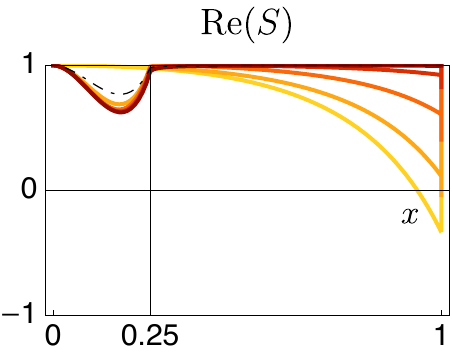}&
                                \includegraphics{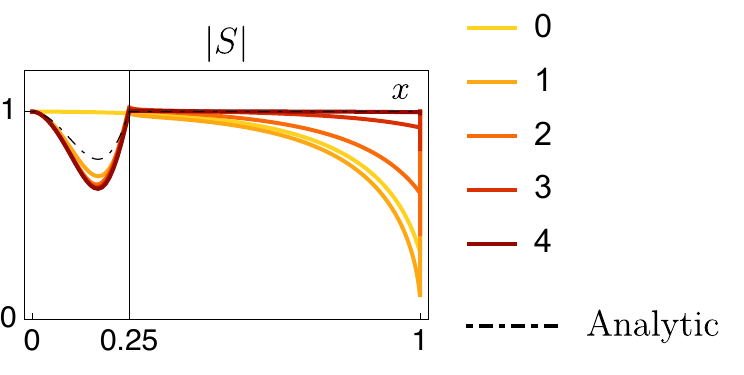}\\
(a) & (b) & (c)                                
 \end{tabular}
 \caption{Results of the algorithm with no bound state and an arbitrary but regular enough starting point (displayed). Convergence is exponentially fast, so we show only the first five iterations. The inelastic input chosen here is given in \eqref{eq:vi-nobs} with $H=120$. Linear interpolants, grid given in \eqref{eq:grid-def}.
   a) Imaginary part of the S-matrix: iterations (color), analytic (dot-dashed). (b) Real part. (c) Modulus of the S-matrix. 7 seconds for 100 iterations and the grid of \eqref{eq:grid-def}.}
  \label{fig:no-bs}
\end{figure}

In figure \ref{fig:no-bs}, we display the results of the iteration for the following inelasticity:
\begin{equation}
  \label{eq:vi-nobs}
  v_i(x)=H x(1-4x) \text{ if }x\leq 1/4,\quad v_i(x)=0 \text{ otherwise}
\end{equation}
(see \eqref{eq:vi-fi} for the relation between $v_i$ and $f_{i}$), starting on an arbitrary starting point.
The parameter $H>0$ reduces to one dimension the space of inelastic inputs. Inelasticity starts at $x=1/4$ which corresponds to the four-particle threshold.

\begin{figure}
  \centering
  \includegraphics{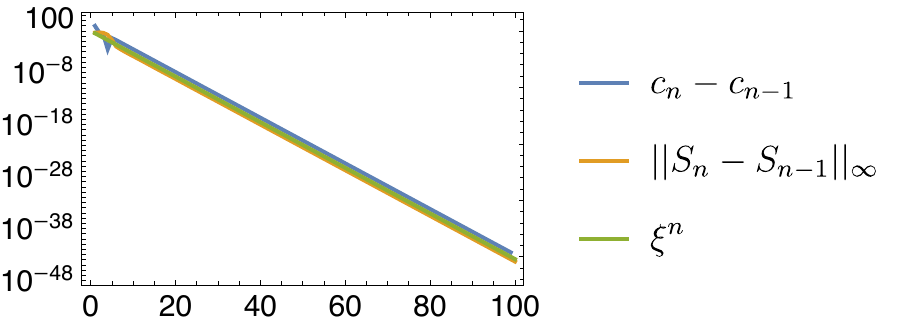}
  \caption{Convergence of the algorithm is dictated by the spectral radius $\xi$ of the Jacobian. The horizontal axis labels the number of iterations. $||.||_\infty$ represents the maximum of the elements of the discretized S-matrix seen as a vector of the values of the S-matrix, $||S_n-S_{n-1}||_\infty = \mathrm{Max}_{i=0,\dots,N}|S_{n,i}-S_{n-1,i}|$ where $S_{n,i}=S_n(x_i)$.}
  \label{fig:pconvlog-no-ms}
\end{figure}

It is remarkable to observe that the speed of convergence is exactly given by the maximum eigenvalue of the Jacobian. This can be seen on figure~\ref{fig:pconvlog-no-ms}.

Convergence is lost when $v_i$ becomes so large that \eqref{eq:finel-def} cannot be satisfied. In our parametrization, this corresponds to $H=197$. 
Near $H=197$, convergence is no more exponential, and eventually leads to divergence, as expected.

A rather wide class of starting points can be used. We tried a variety of functions, but we did not play the game of trying to characterize this space exactly, since the result of the iteration is unique and determined by \eqref{eq:solution}, therefore there cannot be any subtleties in the final answer. In particular, one can start on a null initial amplitude $\rho_0(x)=0$. 

\subsubsection{Examples with one bound state : with and without inelasticity}
\label{sec:one-bound-state}

We now introduce one bound state at mass $s=m_p^2$ ($x=4/m_p^2$) in the analysis. From now on, we use the version of the algorithm where we keep the $c_\infty$ fixed and update the coupling, therefore we have a sequence $g_n$ as well as a sequence $\rho_n(x)$.

We provide two examples to illustrate the dynamics of the algorithm, with, and without inelasticity. Then we discuss the function space in which convergence is achieved below in section~\ref{sec:spectr-radius}.

\paragraph{No inelasticity}

\begin{figure}[h]
    \hspace{-24pt}   
  \centering
  \begin{tabular}{m{5cm}  m{5cm} m{5cm}}
\includegraphics{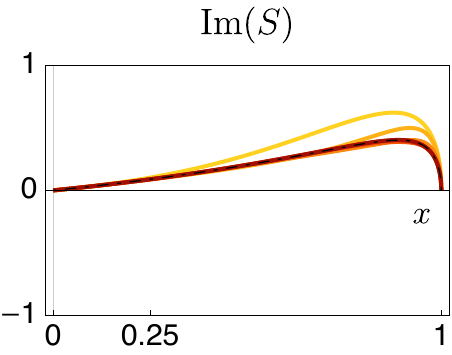} &                   \includegraphics{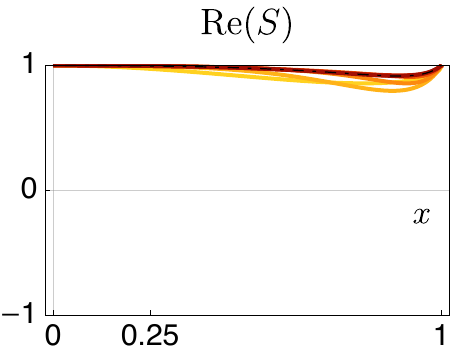}&
                                                                                       \includegraphics{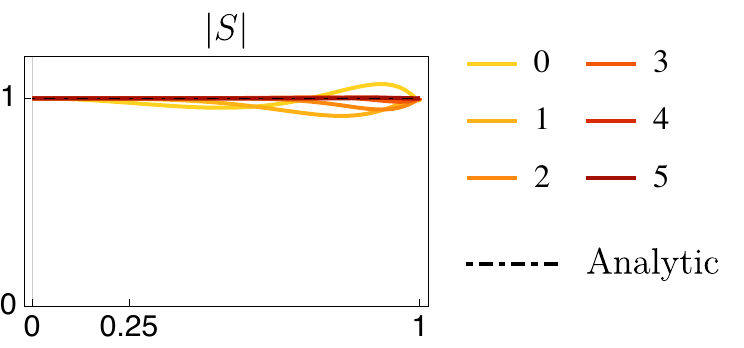} \\
    (a)&(b)&(c)
 \end{tabular}
  \caption{Results of the iterations for one bound state and fixed constant at infinity. Obtained for $c_\infty=1.25$, $m_p^2=2.8$. We show only first 5 iterations.  The analytical solution is given by \eqref{eq:solution} with $S_{\text{elastic}} = S_{\text{CDD}}^{\text{pole}} S_{\text{CDD}}^{\text{zero}}$  with $m_z^2 \simeq3.594...$, the value which solves \eqref{eq:constant} with $v_i=0$. Linear interpolant, grid of eq.\eqref{eq:grid-def}. A hundred iterations are performed in about 5 seconds.}
  \label{fig:pure-bs}
\end{figure}

In the case without inelasticity, the input of the algorithm is solely the mass of the bound state and the constant at infinity. Given this data, and for a wide class of input functions, we found systematically amplitudes with one zero through the fixed point iteration. An example thereof is provided in figure~\ref{fig:pure-bs}.

The algorithm converges to amplitude with one zero, whose location is exactly the one fixed by \eqref{eq:constant}. In this particular example, we found in this example to be $m_z^2 \simeq 3.594...$. Figure \ref{fig:pure-bs} shows the perfect agreement between this solution and the result of the iteration.

With the given input data, other options would have included 2 or more CDD zeros whose contribution to the constant at infinity adds up according to \eqref{eq:locationzero-c-inf}. As alluded to above, many such solutions are possible. 
However, we also observed that for those solutions, the spectral radius is always larger than one, hence, the version of the algorithm considered in this section does not converge. We discuss this in more detail below.

\paragraph{With inelasticity}

In the presence of inelasticity, we similarly found that we converge to the solution \eqref{eq:solution} with only one-zero. 
We show an example of the iteration for the sake of the reader's curiosity in fig.~\ref{fig:bs-inel} below.
\begin{figure}[h]
\hspace{-24pt}
  \centering
  \begin{tabular}{m{5cm}  m{5cm} m{5cm}}
\includegraphics{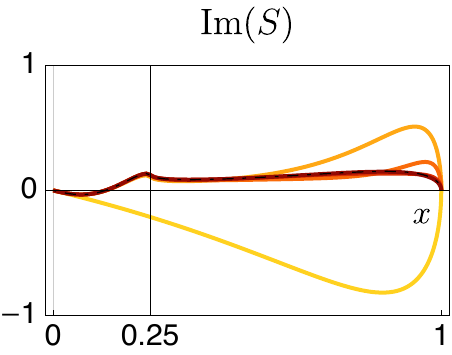}&  
\includegraphics{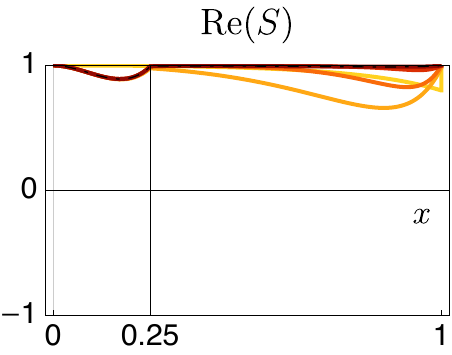}&
\includegraphics{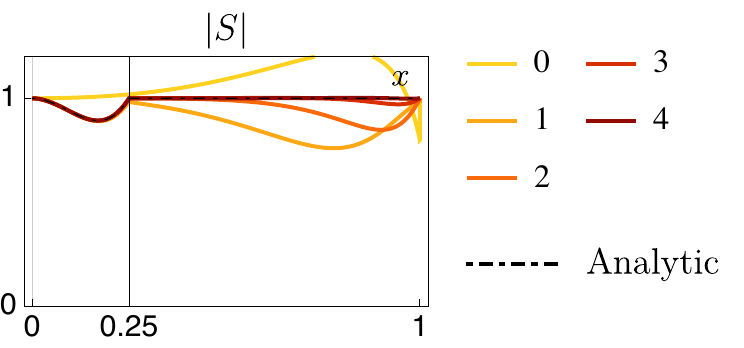} \\
       (a)&(b)&(c)
 \end{tabular}
  \caption{Results of the iterations for one bound state and a fixed constant at infinity. Obtained for $c_\infty=-3$,$m_p^2=2.8$, $H=40$. The result of the first four iterations only. The analytic solution is given by \eqref{eq:solution} with $S_{\text{elastic}} = S_{\text{CDD}}^{\text{pole}} S_{\text{CDD}}^{\text{zero}}$  with $m_z^2=3.225...$, as per eq.\eqref{eq:constant}. Linear interpolant, grid given in eq.~\eqref{eq:grid-def}.}
  \label{fig:bs-inel}
\end{figure}

As we vary some of those parameters (inelasticity when present, or constant at infinity), convergence is lost, and this draws the boundary of some domain of convergence, which we describe at length in section \ref{sec:relat-optim-coupl} below.

\subsubsection{More pairs pole-zero}

Implementing the algorithm for more bound states presents no technical difficulty and can be done easily. 
If we have $n$ bound states, the input data of the algorithm, in the spirit of \cite{Paulos:2016but}, will be the location of the poles $m_{p_1},\dots,m_{p_n}$, the couplings of the poles except the first one $g_2,\dots,g_n$ and the constant at infinity $c_\infty$.
The simplest convergent possibility for this case is the amplitude with $n$ CDD zeros. The problem is therefore to determine $n$ variables: the coupling $g_1$ and the location of the poles $m_{z_1}\dots m_{z_n}$. Each coupling $g_i$ at a given value of the $m_{p_j}$'s furnishes one constraint on the $m_{z_j}$'s, therefore we have $n-1$ constraints coming from the couplings and an extra constraint form the constant at infinity which allows to determine exactly the end-point of the algorithm in advance.

We implemented this algorithm for two poles (linear inerpolant) and observed perfect convergence to the two-zero solutions for a variety of input points, inelasticities, constant at infinity and coupling $g_2$, which lead to conjecture that the systematics is exactly identical. Only the shape of the domain in which convergence is achieved is changed. We have not performed an exhaustive study of this shape in the case with more poles.

\subsection{Spectral radius and convergence of the algorithm}
\label{sec:spectr-radius}

Let us now turn to the analysis of the region in parameter space in which the algorithm converges.

\subsubsection{Odd CDD sector divergence}
\label{sec:odd-cdd-sector}

In the convergent examples above, the total number of CDD factors (poles and zeros) $N_{tot}$ was always even. It is easy to understand the origin of the fact that we never converged to solutions with $N_{tot}$ odd: this is related to the near-threshold behavior \eqref{eq:near-threshold}. When $N_{tot}$ is odd, we have the following near-threshold behavior of the amplitude 
\be
\label{eq:imthres}
N_{tot} ~~~ \text{odd}:~~~ \rho(x) &= 8 \sqrt{1-x} \left( 1 + O(1-x) \right) \ ,  \\
\Re T(x) &= O(1-x) \ . \nn
\ee
With this near threshold behaviour, it is straightforward to show that there exists at least one eigenvector of the Jacobian with the eigenvalue close to $2$ and therefore the iterative map is divergent, as explained below the presentation of the algorithm for the fixed-point iteration, around \eqref{eq:discrete-Jac}.

To demonstrate this explicitly, let us consider the behavior of the Jacobian \eqref{eq:discrete-Jac} near $x=1$. From the behavior \eqref{eq:imthres} and \eqref{eq:discrete-Jac} it immediately follows that
\begin{equation}
  \label{eq:Jac-ev-2}
  J_{N-1,N-1}\simeq 2, ~~~ J_{N-1,i} \simeq 0. 
\end{equation}
This implies that $(J^T)_{i j} \delta_{j N-1} \simeq 2 \delta_{i N-1}$ and therefore $J$ has an eigenvalue close to $2$. As a result even if we start exactly on the solution,  finite grid precision induces a small deviation from the actual solution which grows exponentially and eventually causes the map to diverge after a finite number of iterations (empirically, 5-10). 

\begin{figure}
  \centering
  \begin{tabular}{m{5cm}  m{5cm}}
\includegraphics{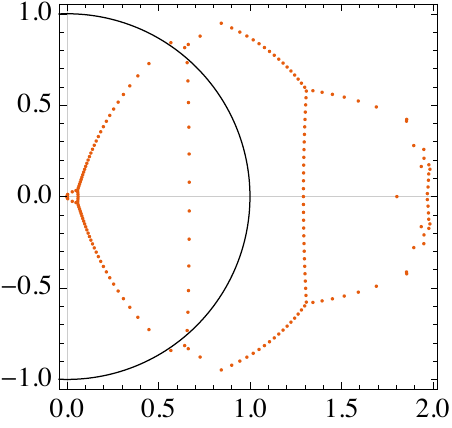}&  
\includegraphics[scale=0.75]{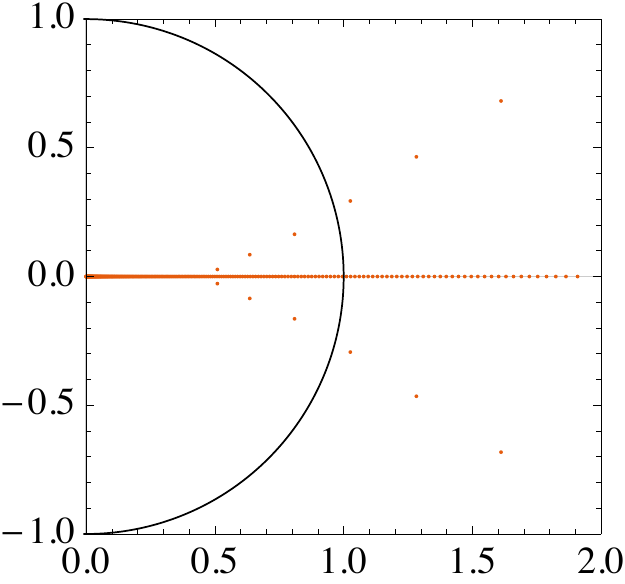} \\
       (a)&(b)
 \end{tabular}
 
  \caption{Eigenvalues of the Jacobian for a pure CDD-pole amplitude for $m_p^2=2.8$: a) linear interpolation; b) interpolation with Bernstein polynomials.}
    \label{fig:ev-atk-CDDodd}
\end{figure}

When we compute the eigenvalues of the iterative map explicitly, we discover an intricate pattern of complex eigenvalues, but confirm the presence of eigenvalues at close to $2$. We show one example in figure \ref{fig:ev-atk-CDDodd}. The e.v.'s are intriguingly contained in a circle of radius one centered around one, but unfortunately the criterion for convergence is that the e.v. should be within the unit disk centered around zero. This might be related to the e.v.'s of Newton's method, but in this case the criterion for convergence is different: as we explain below, the divergence then comes from the fact that the Jacobian becomes singular.

Within the CDD-even class, we have found convergence only when the number of zeros equals the number of poles, and within a certain range of parameters among them, which we discuss in the subsection below. For amplitudes with even number of CDD factors but different numbers of poles and zeros, there seem to exist no region where the iteration converges. We leave understanding this more systematically for the future work.

\subsubsection{Convergence of the 1-pole 1-zero solution. Relation to the optimal coupling analysis}
\label{sec:relat-optim-coupl}

\begin{figure}
  \centering
  \includegraphics{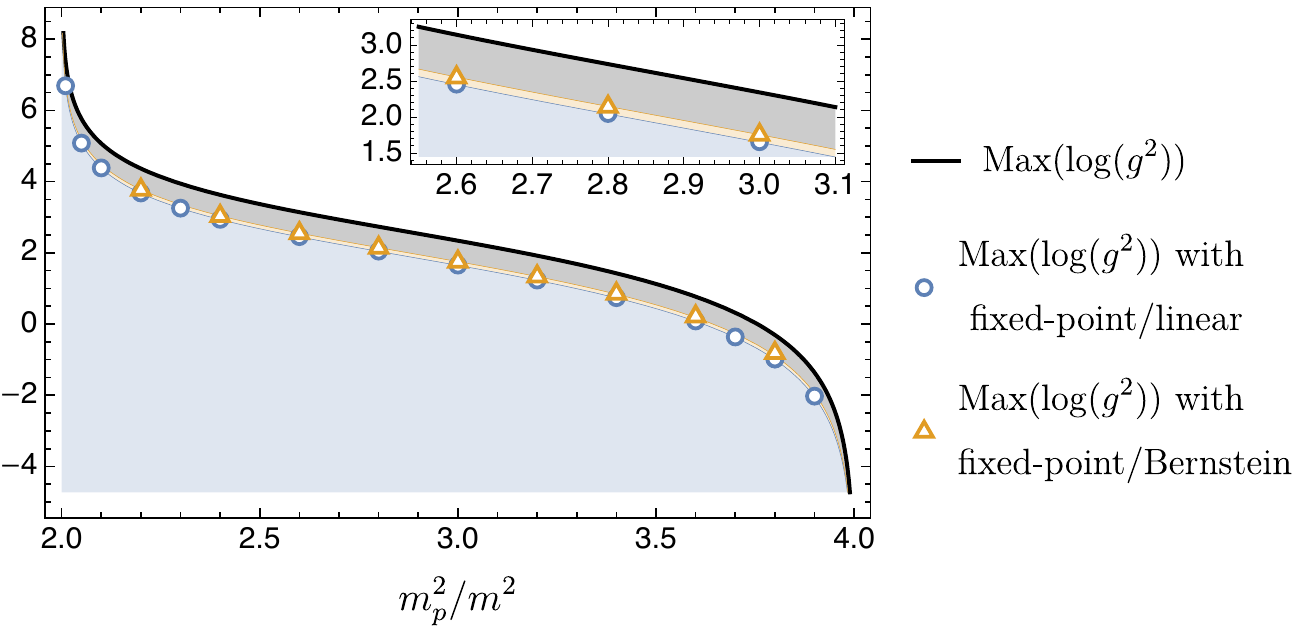}
  \caption{Convergence of the fixed point iteration occurs in the blue-shaded and yellow-shaded regions, depending on the interpolation scheme.}
\label{fig:opt-Atk}
\end{figure}

In this section we present one of the main results of the paper, summarized in figure~\ref{fig:opt-Atk}.

We analyzed the shape of convergence-space for the one-zero one-pole amplitudes. Remarkably, it happens to be given by translates of the optimal coupling curve of \cite{Paulos:2016but}. We also found that, both with linear and Bernstein interpolants, the full space of one-zero one-pole amplitudes cannot be covered and the algorithm stops converging at a small but nonzero distance away from the full space (in section~\ref{sec:newt-kant-meth} we show how Newton's method allowed us to extend this domain).

Taking for granted that our algorithm, when fed with one pole as in \eqref{eq:schem-atk2}, converges only to a 1-pole solution (see hereafter sec.~\ref{sec:non-convergence}), the problem of the shape of the convergence space becomes effectively one-dimensional. Indeed, the inelastic contribution is fixed entirely by \eqref{eq:constant}, and the only unknown that remains is the position of the zero, or equivalently the coupling of the residue of the pole (they are related, hence determining one fixes the other).
As there is only one unknown, we can choose to represent it either as the location of the zero, the value of the coupling, or the value of the constant at infinity, since all these quantities are related to each other.  To allow a more direct comparison with \cite{Paulos:2016but}, we choose to represent the coupling, or rather the logarithm of the coupling squared.

To better understand the results, we need first to describe two phenomena.

Firstly, take a 1-pole 1-zero amplitude. At fixed $m_p$ (mass of the bound-state), the coupling is increased if the CDD zero moves towards $4$, and eventually becomes maximum when the zero is exactly at $4$: that is the optimal coupling amplitude~\cite{Creutz:1973rw,Paulos:2016but} (black curve in figure~\ref{fig:opt-Atk}). That fact is easy to  check by extracting explicitly the residue of the corresponding amplitude. Furthermore, as the constant at infinity is the input of our algorithm and is related to the position of the zero by \eqref{eq:constant} of \eqref{eq:c-zero}, increasing the constant at infinity can also be seen to increase both the zero, and, correspondingly, the coupling, until it reaches its maximal value. Relatedly, when decreasing the constant at infinity, the zero approaches two, and eventually goes off in the complex plane such that $\Re(m_z^2)=2$ and $\overline{m_z}^2 = 4-m_z^2$ so that the constant at infinity remains real nut can then be arbitrarily negative while keeping $\rho(x=0)=0$.\footnote{A complex constant at infinity implies that the imaginary part of the amplitude does not decay to zero and hence the dispersion integral needs to be written with subtractions. We do not consider such amplitudes in this work.}

Secondly, for a fixed constant at infinity, we found (empirically, at first) that increasing the magnitude of  inelasticity increases the coupling. As can be read off the explicit solution \eqref{eq:solution}, the coupling of the solution is given by
\begin{equation}
\label{eq:g-solution}
g=g_{\text{elastic}} e^{\int_{4m^2}^\infty {ds' \over 2 \pi} \log(1 - f_{i}(s')) 
    \sqrt{m_p^2(4 m^2 - m_p^2) \over s'(s'-4 m^2)} 
  \left( {1 \over s' -m_p^2} + {1 \over s' - (4 m^2 - m_p^2) }  \right)} ,
\end{equation}
and the position of the zero is constrained by
\be
c_{\infty} &=2\sqrt{m_{p}^2 (4 m^2 -m_{p}^2)} - 2\sqrt{m_{z}^2 (4 m^2 -m_{z}^2)} \nn \\
&+ \int_{4m^2}^\infty {ds' \over \pi} \log[1 - f_{i}(s')] \left( {1 \over s'(s'-4 m^2)} \right)^{{1 \over 2}} \left( s' - 2 m^2 \right) .
\label{eq:c-zero}
\ee
In \eqref{eq:c-zero}, we see that increasing inelasticity shifts the zero towards 4, hence increases the factor $g_{\text{elastic}}$ above. To determine whether $g$ increases or not, one needs to look at the exponential. It turns out that for most inelasticities, the exponent is numerically very small because of the fast decay ${1 \over s^2}$ of the integrand, and that integral is hardly different from zero. \textit{Therefore, from this perspective, increasing inelasticity increases the coupling.}

We can now describe our results. Our first result is that we mapped the domain of convergence space without inelasticity by applying the process described above: at a fixed value of $m_p$, for which values of the constant at infinity does the map converge? As expected, we found that  the constant at infinity can be arbitrarily negative, and this simply pushes the zero in the complex plane with real part $2$. We find that the actual limit of convergence is reached for \textit{positive values} of the constant at infinity, which are such that the coupling of the resulting amplitude is a certain fraction of the optimal coupling;
\begin{equation}
  \label{eq:opt-coupl}
  {g^2\over g^2_{\mathrm{max}}}\sim c\,.
\end{equation}
With linear interpolants, we found that fraction ${g^2 \over g^2_{\mathrm{max}}}$ to be of order $0.5$, while with Bernstein polynomials we had a slight improvement and observed a factor of $0.55$. This is the blue and yellow thin curves in figure~\ref{fig:opt-Atk}.

We also explored the convergence space in the presence of inelasticity; this is our second result. Surprisingly, we found that inelasticity \textit{does not} play a role, and whatever function we used, we could only reach the same maximal coupling as without inelasticity. Later, when exploring Newton's method in the next section we will again find that remarkably the convergence region is well characterized by the criterion ${g^2 \over g^2_{\mathrm{max}}}$ is less than some number.

Furthermore, we have not been able to produce any improvement of these bounds by improving our grid and we believe that the bound is strict: in two dimensions
the fixed-point iteration map should not cover all of parameter space. In the next section we will increase the region of convergence using Newton's method.

To produce our results, we applied the bisection method by hand for a given grid of mass-values $m_p$: we could in this way found an optimal $c_\infty$ where the algorithm is at the edge of not converging anymore. Our criteria was that, after a few hundred iterations, the S-matrix should be of modulus one to $10^{-3}$ accuracy over the last 10 iterations. We came up with this criterion by empirical search. It turned out to correspond pretty accurately the fact that the spectral radius becomes equal to $1$ (within $10^{-2}$). 

\subsubsection{Non-convergence of $1$ pole-$n>1$ zeros amplitude}
\label{sec:non-convergence}

We now look at amplitudes with one pole and more zeros and ask whether our fixed-point iteration could converge to them.

The first observation is that one can always take an amplitude with $n$ zeros and send $n-1$ of them towards $4$, in such a way that, up to our numerical precision, such an amplitude becomes undistinguishable from the the $1$-zero amplitude. This is the sort of CDD ambiguity that we have nothing to say about. For instance, by setting the zero to its maximum value for the 1-pole 1-zero problem, we observe that, within our numerical accuracy, a second zero should be located at $3.99999$ i.e. $10^{-5}$ away from $4$, in order to have a spectral radius below one and hence possible convergence. At this point, the S-matrices look completely identical and this relates to the question of the numerical accuracy of the algorithm.

Excluding this pathological behaviour, we have scanned extensively the bulk of parameter space of the 1-pole $n$-zero amplitude (n even and odd) up to $n=5$.

We first designed a grid of $n$-tuples $(m_p,m_{z_1},\dots,m_{z_n})$ of 10 evenly-spaced, \textit{distinct} elements within $[2.1;3.9]$ (11 for $n=5$). We computed systematically the spectral radius of the Jacobian for all the elements (for $n=5$, there are more than 400 tuples) and found all spectral radii to be equal exactly to two, to $10^{-2}$ accuracy for linear interpolants. We also computed systematically for $n=2$ zeros the spectral radius for possibly identical elements (which include zeros of higher order). This represents around 1000 eigenvalue computations, which all ended up being equal to $2$ to the same accuracy.

Note that while we presented an argument for existence of such an eigenvalue for odd number of poles and zeros around \eqref{eq:Jac-ev-2}, we do not have such an argument for $N_{tot}$ even. It would be interesting to understand the origin of this fact.

Given the observed similarity of eigenvalue patterns between even and odd $n$, and knowing that for $n$ odd we can determine that the spectral radius is at least two, we are lead to conjecture for all $n$, the spectral radius of the Jacobian of the map of the $1$-pole $n$-zero amplitude is exactly $2$. It would be very interesting to have a demonstration of this fact.

\section{Newton's method}
\label{sec:newt-kant-meth}

In the previous section we explored fixed-point mapping \eqref{eq:fixedpointiter} to solve unitarity and crossing. We observed convergence of the algorithm for a wide class of amplitudes and within some range of parameters (positions of poles and zeros, constant at infinity). We gave evidence that the fixed point iteration diverges at a finite distance away from the boundary of amplitude space, and explained that improving the grid did not decrease this distance.

The aim of this section is to present results obtained using a different iteration scheme, with better convergence properties: Newton's method \eqref{eq:NKmethod}.

In one dimension, Newton's method aims at finding a root of a function $f(x)$ via the following sequence
\begin{equation}
  \label{eq:newton}
  x_{n+1}=x_{n}-\frac{f(x_{n})}{f'(x_{n})} \ .
\end{equation}
For a function continuously differentiable near a root, there always exists a neighborhood of the root such that Newton's method will converge to the root. In this case, convergence will be ``quadratic''.\footnote{This means that the error at one step is proportional to the square of the error at the step before. This is faster than fixed point where convergence is only ``linear'', which means that the error is proportional to the error at the previous step, and typically decreases like $q^n$ for $q<1$.}
Convergence of the Newton method is a rich mathematical domain. One famous result that maybe illustrates this point best is the fractal structure of the basins of attraction of the method to determine complex roots of polynomials, see appendix \ref{sec:fractal-structure} for details. In this section we will observe similar patterns, which we relate to CDD ambiguities, and discuss them in section~\ref{sec:cdd-dis-ambiguation}. 

The point that interests us here is that Newton's method can be used to solve a fixed point equation  $x^*=g(x^*)$, by searching for the roots of $f=\mathrm{id}-g$, where $\mathrm{id}$ is the identity function, $\mathrm{id}(x)=x$. 
Atkinson suggested, see e.g. \cite{Atkinson:1970zz}, that when the  iteration method described in the previous section reaches the boundary of  convergence, one could extend convergence by using Newton's method.
The conditions for convergence of the Newton method are indeed better than the fixed point iteration, because the gradient helps to direct the iteration.

The steps of the iteration as we implemented it are as follows. The steps 1-3: discretization, interpolation, and dispersion integral, are implemented in the same way as before. At step 4 the algorithm changes, as follows:
\begin{itemize}
\item[4'.] {\it Newton's method iteration.} The map \eqref{eq:NKmethod} is implemented as
\begin{equation}
  \label{eq:NK-def}
J^{\Psi}_{n} (\rho_{n+1} - \rho_{n})=- \Psi(\rho_{n})
\end{equation}
where $i,j$ indices have been suppressed, $\Psi \equiv \mathrm{id}-\Phi$, $\mathrm{id}$ is the identity operator, $\Phi$ is the iterative map defined in the previous section in \eqref{eq:map-discrete}, 
$J^{\Psi}_{n,ij}$ is the Jacobian of the map $\Psi$. In effect, this way of implementing Newton's method is equivalent the form 
\begin{equation}
  \label{eq:NK-def-inverse}
  \rho_{n+1} = \rho_{n} - ({J^{\Psi}}_{n})^{-1} \Psi(\rho_{n}) \,,
\end{equation}
whose continuous version was eq.~\eqref{eq:NK}.
But eq.\eqref{eq:NK-def} has an immense advantage in terms of its numerical cost and stability.
\footnote{Linear solvers perform better than brute-force inversion for a variety of reasons. In our case, the main reason seems to be that inversion is tantamount to having solved the system for all values of the input vectors.}
The Jacobian is defined explicitly by
\begin{equation}
  \label{eq:jac-NK}
  J^{\Psi}_{n,ij}=\partial (\rho_{n,i}-\Phi[\rho_{n,i}])/\partial \rho_{n,j} = \delta_{ij} - J_{n,ij}
\end{equation}
where $J$ was defined in \eqref{eq:discrete-Jac}.

In Mathematica, we use $\texttt{LinearSolve}[J_n',b]$, where $b= J_n'\rho_{n} - \Psi(\rho_{n})$, which performs very well in terms of speed for grids of size $\lesssim 1000$ which we worked with (less than a second, where inversion would take of the order of minutes).
\end{itemize}

We present now the results we obtained with Newton's method. Firstly we describe the extension of convergence range in the one-pole one-zero sector, see figure~\ref{fig:opt-NK}. Next we describe the phenomenon of CDD ambiguity, which arises because Newton's method allows convergence in a range where many solutions are possible given the input data, and the end-point of the iteration is fixed by the starting point. We find that the starting points hint to a standard fractal structure of basins of attraction which we describe in figure~\ref{fig:fractal-CDD}.

\subsection{Improved convergence for the 1-pole 1-zero solution}
\label{sec:large-impr-conv}

We first consider the case discussed in detail in section~\ref{sec:relat-optim-coupl}. The S-matrix of interest has one CDD pole and one CDD zero.  When doing the fixed-point iteration, we observed that the algorithm stopped converging as the CDD zero was approaching the two-particle threshold and the convergence was lost when an eigenvalue of the map became bigger than one. Let us now show the results of our analysis using the Newton's method.

\begin{figure}
  \centering
  \includegraphics{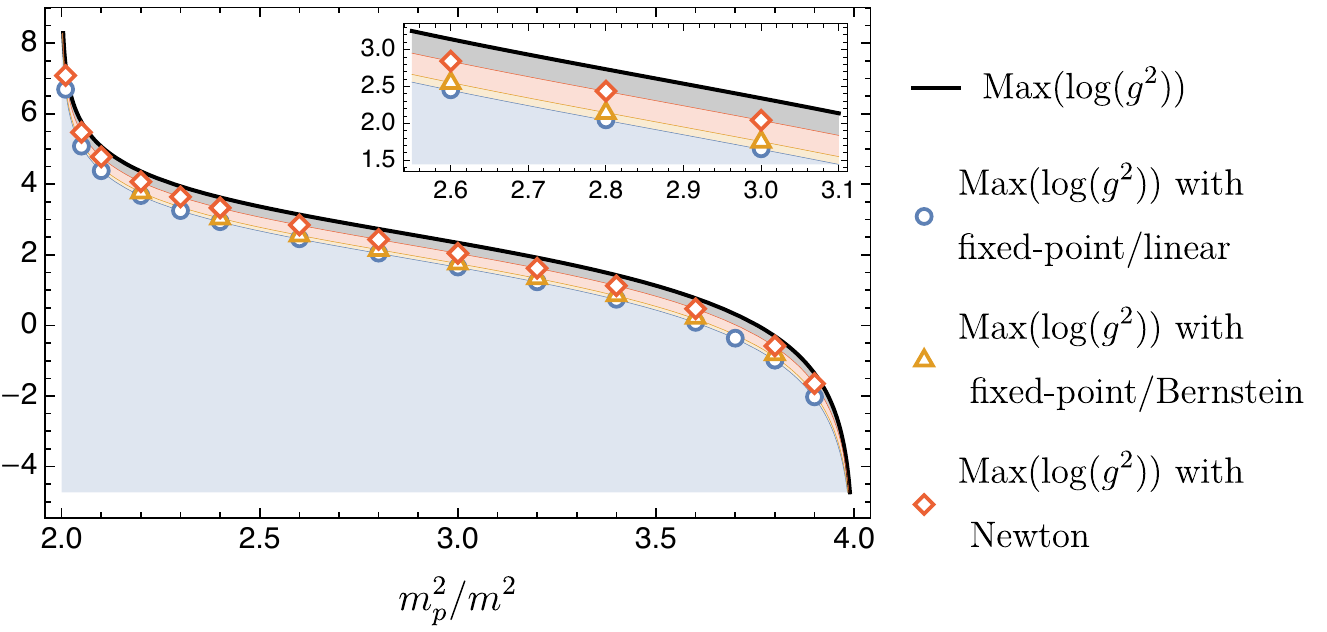}
  \caption{Maximal coupling as a function of the mass of the bound state. The fixed-point iteration method converges in blue-shaded and yellow-shaded regions. The Newton method extends convergence to the red-shaded region. With finite grids that we used we were not be able to fill in the gray region.}
\label{fig:opt-NK}
\end{figure}

We found that Newton's method can be used to significantly extend the region of convergence of the fixed-point mapping. This is an important result from two points of view. From the Atkinson program's perspective, it remained an unknown whether the convergence radius of the fixed-point iteration could be extended at all. Our succesful implementation of Newton's method shows that it is indeed the case. Secondly, from the point of view of our bootstraping procedure, it is important to be able to describe a wider class of amplitudes.

We depict our extended convergence region by plotting again the maximal coupling as a function of the mass of the bound state in figure \ref{fig:opt-NK}.
We could get, with both linear and Bernstein interpolants, to ${g^2 \over g_{\mathrm{max}}^2}\sim  0.75$, which is a significant improvement of ${g^2 \over g_{\mathrm{max}}^2}\sim 0.5-0.55$, which we could achieve using the fixed-point iteration. Interestingly, linear  interpolation gave better results than Bernstein's this time.

\begin{figure}
  \centering
  \includegraphics[scale=1.25]{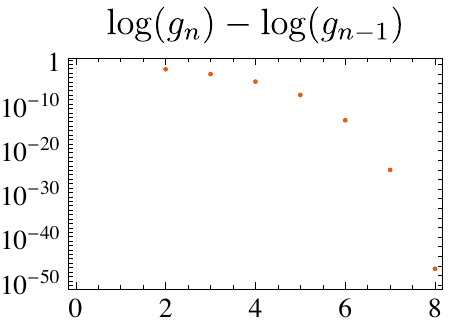}
  \caption{Fast convergence of Newton's method. On the vertical axis we plot the difference between consecutive axis. The horizontal axis labels the number of iterations. To be compared with figure \ref{fig:pconvlog-no-ms}.}
  \label{fig:fast-newton}
\end{figure}

In the bulk of the convergence region, we observe (see figure~\ref{fig:fast-newton}) that the convergence speed of the Newton method is extremely fast, as expected from the fact that it typically converges quadratically: we converge to $10^{-60}$ in 5-10 iterations in the bulk of convergence range. This counterbalances the time lost in computing the Jacobian.

A natural question one then can ask is the following: as we improve the resolution of the grid can we fill in the gray region on figure \ref{fig:opt-NK}? To our surprise, given the efficiency of Newton's method, we found that for a few grid resolutions that we used, we could not improve  the results in a clearcut way. This {would} lead us to {conclude} that there {is} a finite set of amplitudes which the default implementation of Newton's method {is not} able to capture. We motivate this intuition as follows.
\begin{figure}
  \centering
  \includegraphics{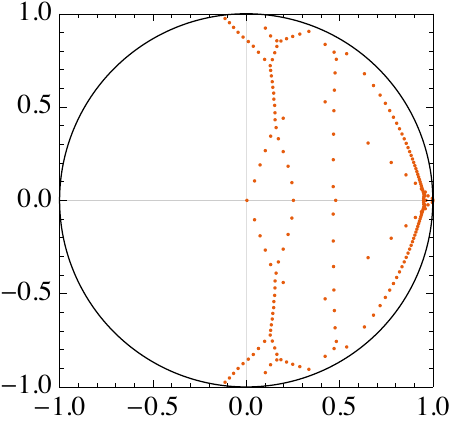}\qquad
  \includegraphics[scale=0.725]{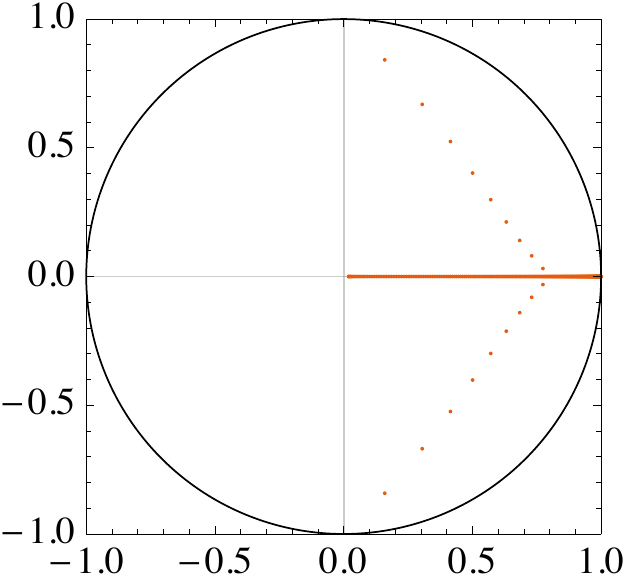}
  \caption{Distribution of eigenvalues of $J^{\Psi}$ for a pure elastic scattering amplitude consisting of a singe CDD-pole and a CDD-zero. We plot it close to the boundary of convergence, here $m_p^2 = 2.4$ and $m_z^2=3.98$ with linear interpolation (left), while $m_z^2 = 3.97$ with Bernstein polynomial interpolation (right). As we move the location of zero closer to $4$, the support of distribution of zeros on the right panel moves to negative values for interpolation with Bernstein polynomials. It has a more obscure pattern for linear interpolants. The amplitude functions, past the limit of convergence, look more and more jagged, with an instability growing near $x=0.9$, as in figure~\ref{fig:instarho}.}

\label{fig:eigenvalues-Bernstein}
\end{figure}
In analyzing how convergence is lost for one-zero one-pole amplitudes, we found that, as $m_z^2$ approaches $4$ (i.e. as we increase the coupling), the Jacobian $J^{\Psi}$ becomes singular and develops a zero eigenvalue. We plot the distribution of eigenvalues for $J^{\Psi}$ close to the boundary of convergence in figure \ref{fig:eigenvalues-Bernstein}.

We observed that the distribution of eigenvalues does not qualitatively change as we change the grid. It simply becomes more dense along the lines that are clearly visible on both plots of figure~\ref{fig:eigenvalues-Bernstein}. In particular, for Bernstein's interpolants, the picture is very clear: the e.v.'s populate a continuum on the real line and once the minimal e.v. has passed $0$, the Jacobian becomes uniformly singular. The situation is a little more complex for linear interpolants, where one could think that if a line of e.v.'s crosses the origin, the algorithm might converge again. We observed that this is not the case and passed this point, the algorithm either diverges or yield pathological results.

\begin{figure}
  \centering
  \includegraphics[scale=1]{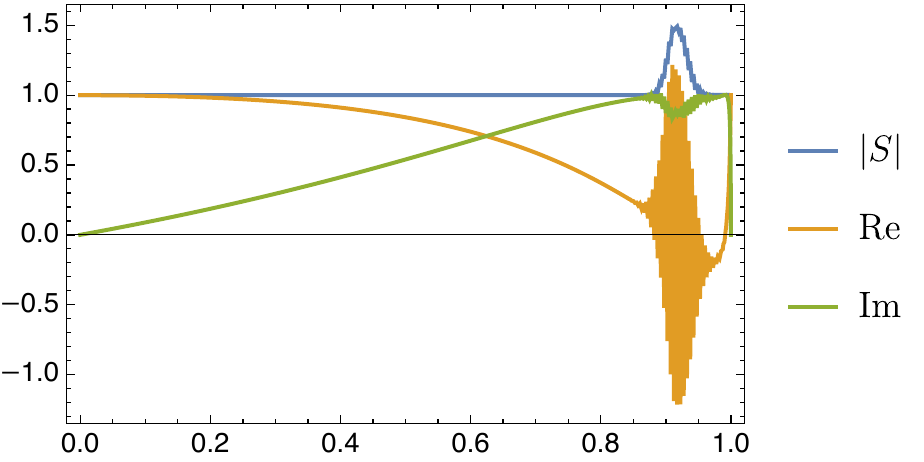}
  \includegraphics[scale=1]{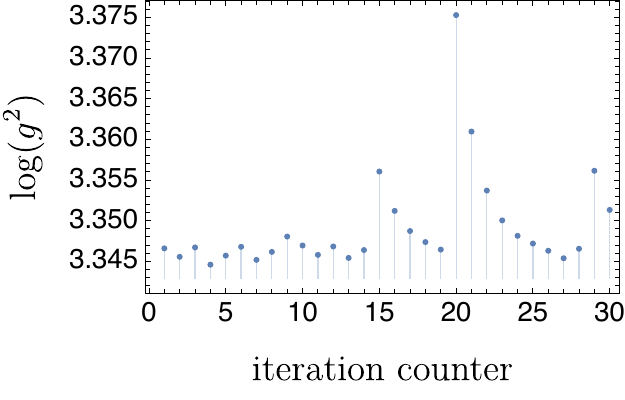}
  \caption{An example of unstable behavior of Newton's method with linear interpolations for $m_z^2=3.981$ after 30 iterations.}
  \label{fig:instarho}
\end{figure}

We therefore believe that, in order to fill the gray region that remains in figure \ref{fig:opt-NK}, a more sophisticated method of solving unitarity is needed. It would be interesting to develop such a method, but we do not pursue this further in the present paper. For illustrative purposes, in figure~\ref{fig:instarho} we display an example of non-convergent iteration past the convergence limit for linear interpolants. The functions exhibit an instability growing near $x=1$ and look more and more jagged or dented, while on the coupling plot one can see oscillations appearing.

\subsection{Accessing different CDD sectors and a CDD fractal}
\label{sec:cdd-dis-ambiguation}

Given that now we are in possession of an algorithm that converges for a much broader set of amplitudes, we can ask again the question which was considered in section~\ref{sec:conv-one-bound}. Given an input data which is: the constant at infinity and the position of the poles, how does the algorithm determines the number of CDD zeros in the final solution? Indeed, many combinations of CDD-zero factors can produce the same constant at infinity, and it cannot be known in advance what the algorithm converges to. This is one aspect of the CDD ambiguity.

\begin{figure}
  \centering
  (a) \includegraphics[scale=0.48]{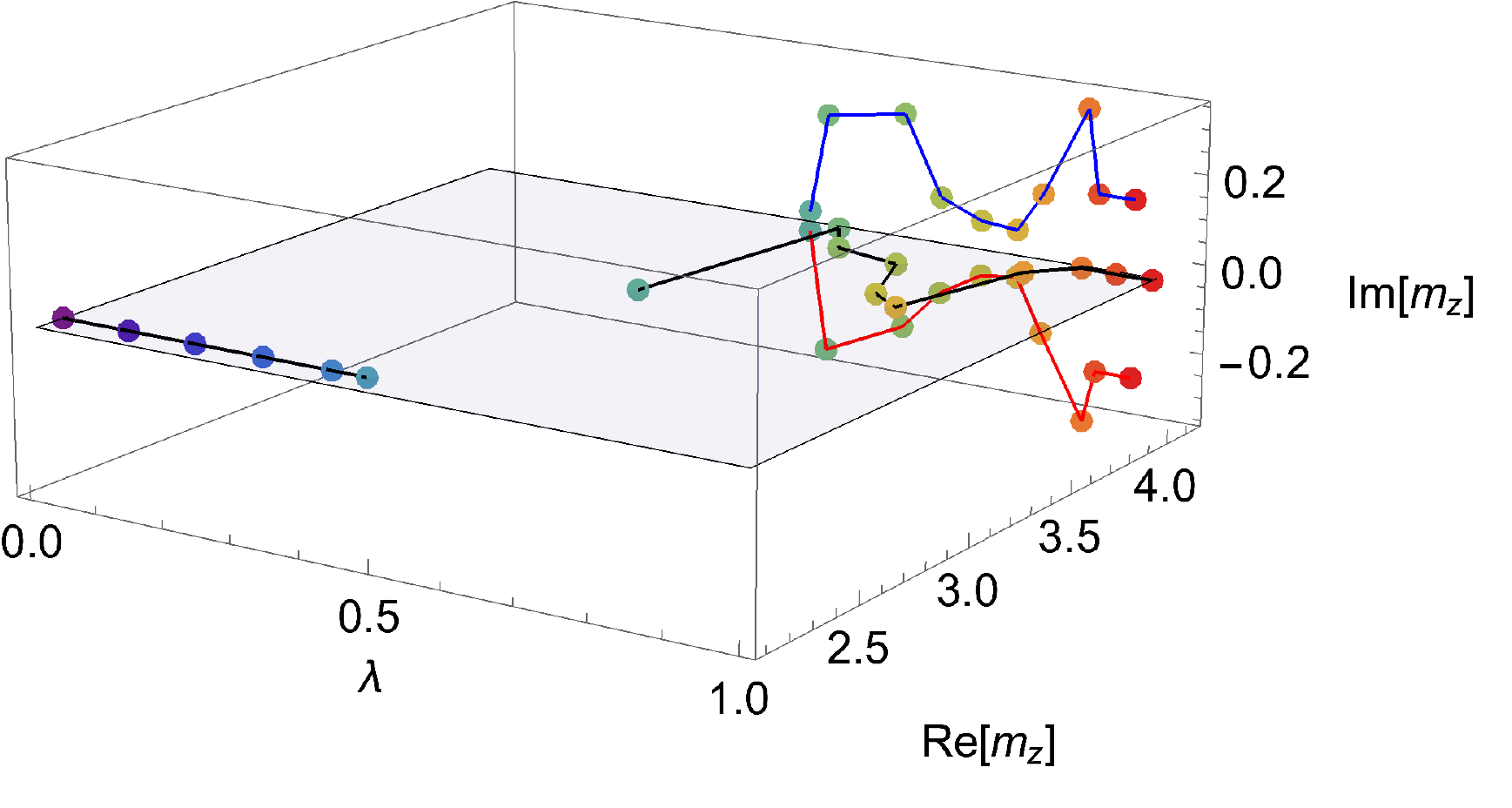}\quad
  (b) \includegraphics{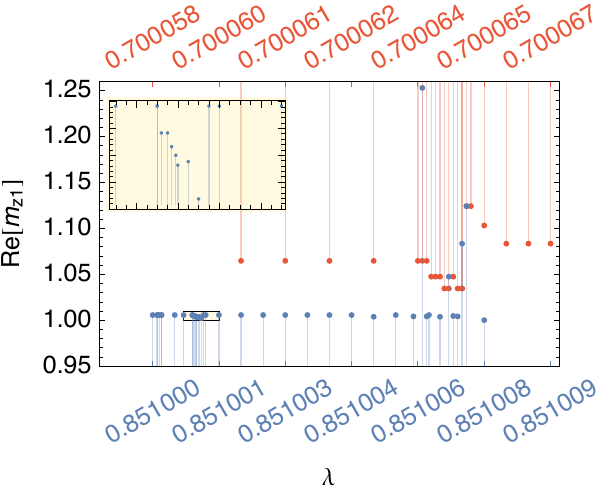}\\
  \vspace{12pt}
  (c)  \includegraphics[scale=0.32]{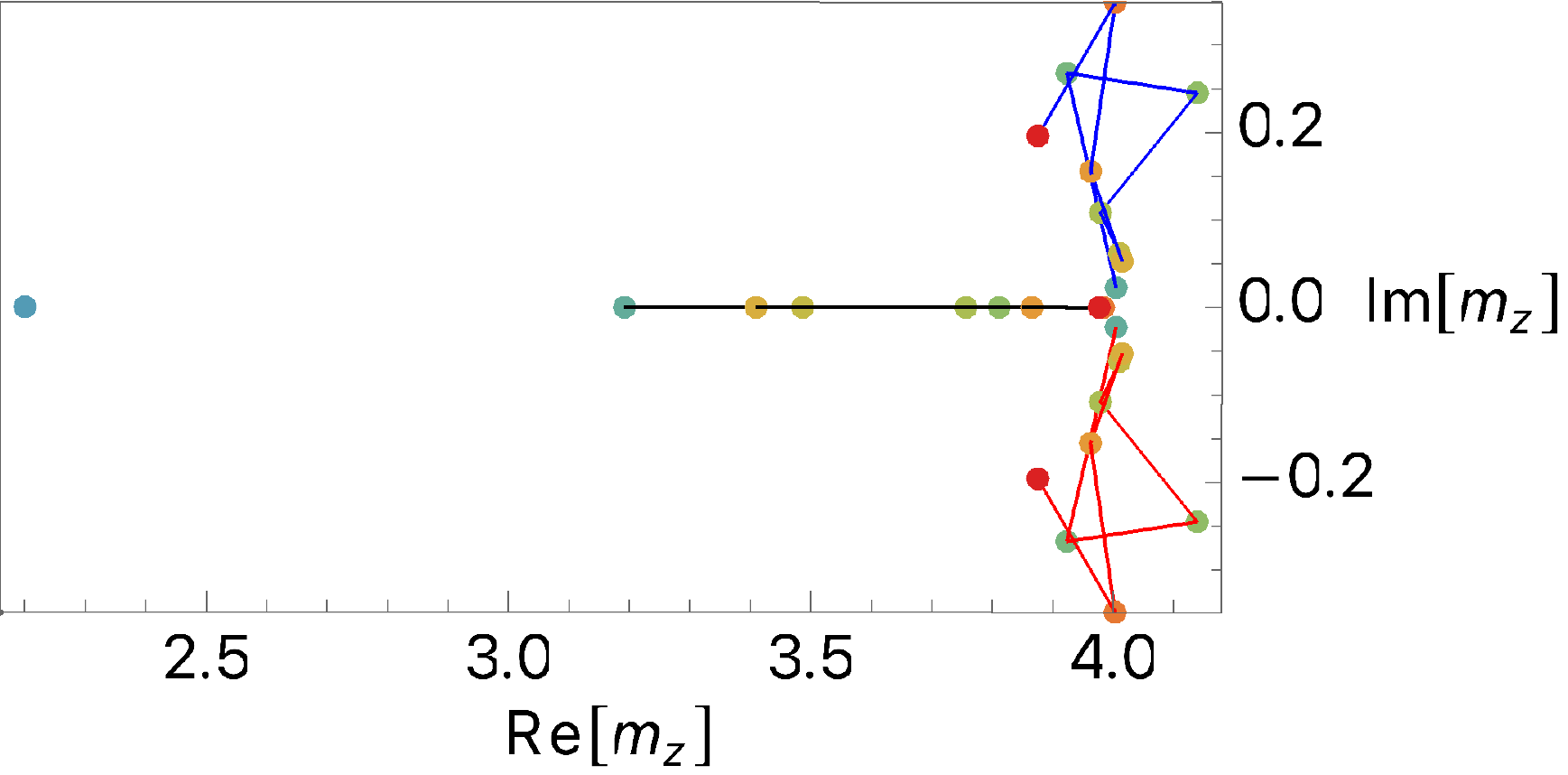}\quad
 (d) \includegraphics[scale=0.32]{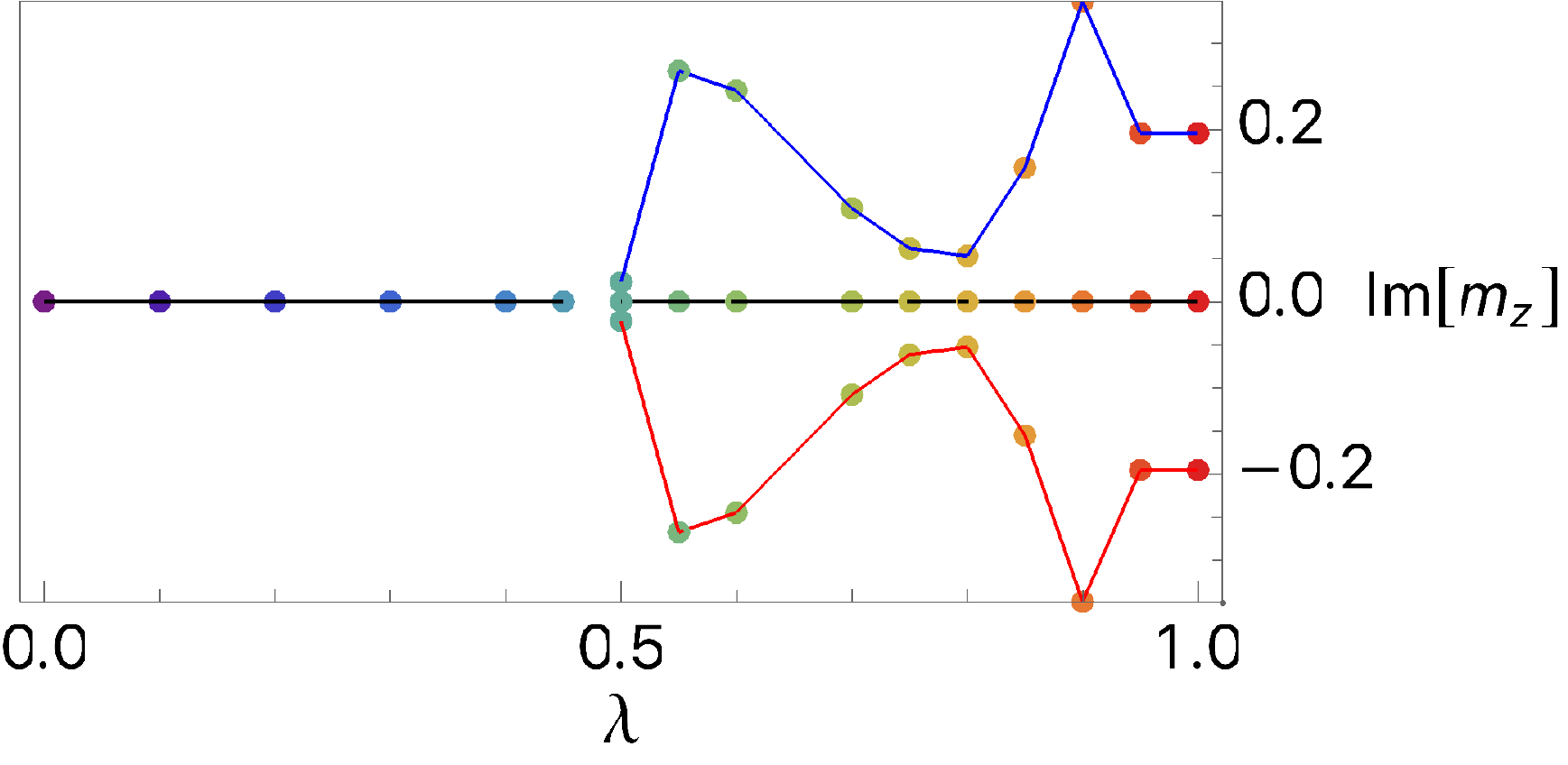}\quad
  \includegraphics[scale=0.8]{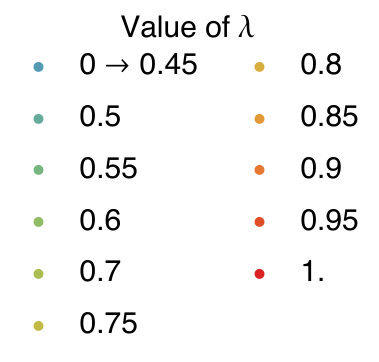}
  \caption{CDD fractal. (a) Position of the zero(s) of the amplitude as the initial point, determined by $\lambda$ in eq.~\eqref{eq:T0lambda} is varied. For $\lambda<1/2$ we have the one-zero amplitude, and for $\lambda\geq1/2$ we find three zeros which move in the complex plane. (b) Shows the real part of the real zero $m_{z_1}=\Re(m_{z_1})$, as a function of $\lambda$, on two superposed $x$-axis (top and bottom for ticks). Panels (c) and (d) represent projections of the three-dimensional plot a). Those plots illustrate the roughness of the curve for $\lambda\geq1/2$, at various scales, around $\lambda=0.7$ (red) and $\lambda=0.851$ (blue). This is indicative of some chaotic fractal behaviour. Hence the name: CDD fractal. }
  \label{fig:fractal-CDD}
\end{figure}

The answer to this question is actually simple: it depends on the starting point. We have been able to observe this very explicitly. By choosing generic initial conditions we found that the algorithm converges to the one pole and one zero solution. By tuning the initial data close to another many-zero solution, we could reach the solutions with multiple zeros as we demonstrate in figure \ref{fig:fractal-CDD}. 

In this figure, we report the results of the following experiment. We chose a particular CDD amplitude with one pole at $m_p^2=2.1$ and one zero at $m_z^2=2.2$. We further chose one CDD amplitude with the same pole and three zeros at $m_{z_1}^2=3.8,\,m_{z_2}^2=3.9$ and the last one $m_{z_3}^2$ such that the constant at infinity of both amplitudes are identical, which means that $m_{z_3}^2$ solves the following equation:
\begin{equation}
  \label{eq:mz3}
  \sqrt{m_{z_1}^2(4-m_{z_1}^2)}+  \sqrt{m_{z_2}^2(4-m_{z_2}^2)}+  \sqrt{m_{z_3}^2(4-m_{z_3}^2)}=  \sqrt{m_{z}^2(4-m_{z}^2)}
\end{equation}
Numerically, this gives $m_{z_3}^2\simeq 3.938$. 

We then considered a series of starting points of the iteration algorithm $\Im T_0(x) = f_\lambda(x)$ that interpolates between the one-zero amplitude and the three-zero amplitude defined by
\begin{equation}
  \label{eq:T0lambda}
  f_\lambda(x) =(1-\lambda) \Im T_{1-{\rm zero}}(x,m_p,m_z)+\lambda \Im T_{3-{\rm zero}}(x,m_p,m_{z_1}, m_{z_2}, m_{z_3}),\quad \lambda\in[0;1]
\end{equation}
We asked the following question: how does the result of Newton's method iterations depend on $\lambda$?

In the first part of our analysis, we studied a uniform grid spacing in $\lambda$-space with increments of $0.05$. We discovered that for $\lambda<1/2$, we systematically converge to the one-zero solution. We probed more values in this range, in particular close to $\lambda\lesssim1/2$, and confirmed this behaviour.
With linear interpolants, the first signs of different amplitudes arise very close to $\lambda=1/2$ where for instance at $\lambda=0.499992$ we have a one-zero solution, $\lambda=0.499993$ displays a 7-zero solution, while $\lambda=0.499994$ is just a 3-zero solution. With Bernstein interpolants, the transition 
happens just above $0.5$, but the rough behaviours are identical. Then, for other evenly spaced values of $\lambda\geq1/2$, we observed that this somewhat rough behaviour persists. The amplitude converges most of the time to a 3-zero amplitude, whose position of the zeros in the complex plane we have represented as a function of $\lambda$ in 3 dimensions in figure~(\ref{fig:fractal-CDD}a). The position on the real zero on the blue-shaded plane in particular is visibly erratic on this plot. We did not represent this but the complex zeros also undergo such an erratic movement in the complex plane.

The abrupt change around $1/2$ and the appearance of very different amplitudes within a small range of $\lambda$ pushed us to repeat the experiment near other points of $\lambda>1/2$ but with a greater resolution (we did this with linear interpolants). We chose $\lambda=0.7$ and $\lambda=0.85$ for no special reason. We observed at various scales a similar roughness, as we try to illustrate in figure~(\ref{fig:fractal-CDD}b) where we represented only the position of the real part of the real zero. This figure shows the two series of points $\lambda=0.7$ and $\lambda=0.85$ in red and blue, respectively, whose x-graduations are read on the top and bottom axis, respectively. A zoomed version of a transition near $0.8510005$ is shown as a yellow-background subgraph. We also found amplitudes with more zeros (5) at other selected points. Overall, this is strongly indicative of a chaotic fractal behaviour: within flat basins of attraction of some various sizes, sudden transitions occur. It is very similar to the fractal structures classically found using Newton's method when finding roots of one-dimensional function, see appendix~\ref{sec:fractal-structure}.
It is therefore not surprising to find a similar behaviour here, in the context of the CDD ambiguity, which one could refer to as a ``CDD fractal''. We did not try to characterize this phenomenon in a more detailed way. It would be interesting to see if a similar phenomenon occurs for higher-dimensional amplitudes as well.

As we take the continuum limit, several scenarios are possible. It might be that the size of all of the flat basins of attraction shrinks to zero and give a nowhere-continuous curve. 
It might also be that finite size basins of attraction remain at all scales. It is an interesting open question.

Finally, using Newton's method, we were able to obtain solutions with odd $N_{tot}$. This is remarkable because this was not possible using the fixed-point iteration method. This fact also excludes the possibility that the square-root behaviour near $s=4$ presents a problem for the algorithm of the algorithm, because all amplitudes with odd $N_{tot}$ exhibit the same near-threshold behavior \eqref{eq:imthres}, and not just the pure CDD-pole amplitude for which Newton's method does not converge.

\section{Discussion and future directions}
\label{sec:discussion}

\subsection{Discussion of the different strategies}
\label{sec:disc-diff-strat}

We used two different iterative strategies to solve equation \eqref{eq:atk-schem} together with analyticity and crossing: fixed-point method \eqref{eq:fixedpointiter},  and Newton's method \eqref{eq:NKmethod}. We further used two different interpolating strategies for each of the algorithms: linear interpolation described in appendix \ref{sec:linear-interpolation}, and interpolation with Bernstein polynomials described in appendix \ref{app:bernstein}. Here we review the pros and cons of various strategies.

\subsubsection{Fixed-point \textit{vs} Newton}
\label{sec:fixed-point-vs}

To remind the reader, the iterative map was defined by
\begin{equation}
  \rho_{n+1}=\Phi[\rho_n]
\end{equation}
and Newton's method was applied to find the roots of $\Psi \equiv \text{id} -\Phi$
\begin{equation}
  \rho_{n+1}=\rho_n-(\Psi')^{-1} \cdot \Psi [\rho_n] \ .
\end{equation}
While both methods lead to the same result, namely a solution to unitarity and crossing, the domains in which the algorithms converge are not identical.

We could partially characterize domains of convergence in both cases. For the fixed-point iteration we related convergence of the algorithm to the spectral radius (the maximum eigenvalue, in modulus) of the Jacobian of the $\Phi$ \eqref{eq:discrete-Jac}.  The fixed-point iteration converges whenever the spectral radius at the fixed-point is smaller than one. For Newton's method, convergence is lost whenever the Jacobian of $\Psi$ becomes singular. There exist adaptions of the Newton method which can be employed to deal with singular Jacobians. We did not attempt this: since the problem of unitarity and crossing in two dimensions is already solved theoretically, improving this aspect did not seem pressing.

Newton's method is also slightly more complex to implement, as one needs to compute the Jacobian of the map at every step. Since it was easily computable for us, it is not surprising that we could implement this method efficiently. If the Jacobian is not available or hard to compute, one can use the so-called quasi-Newton methods. Note also that the time spent in computing the Jacobian is a trade-off for much faster convergence of Newton's method.

\paragraph{Results}
We observed that implementation of Newton's method increased vastly the range of convergence of the fixed-point iterative algorithm. It extended convergence in two directions. Firstly, it allowed to describe more 1-zero 1-pole amplitudes, see figure \ref{fig:opt-NK}. Secondly, it allowed to describe amplitudes with arbitrary number of CDD zeros. In the latter case, we observed a rich pattern of convergence to various solutions as a function of the starting point, with a fractal structure, see figure~\ref{fig:fractal-CDD}.

Newton's method also converged much faster than the fixed-point iteration, due to the help of the gradient. Compare for instance figures \ref{fig:pconvlog-no-ms} and \ref{fig:fast-newton}.
From Atkinson's program perspective, these results are truly new, as it had not been shown, even theoretically, that one could go beyond the convergence radius of the fixed-point iteration. Our results show that it can be done.

\subsubsection{Linear interpolation \textit{vs} Bernstein polynomials}
\label{sec:linear-vs-bernstein}

We used two different types of interpolants : linear, and polynomial (Bernstein). For both, we were able to devise grids that would yield results with good accuracy, within a certain range of parameters, with laptop computing power in \texttt{Mathematica}.
The results for finite grid-size, which we obtained were very good and in principle we could converge to an even better accuracy by increasing the size of the grid (see below).

The advantage of the linear interpolants is that we were able to locate more points near $x=1$ ($s\to4^+$). It was however a delicate task to know where specifically to add points to improve convergence, while with Bernstein polynomials is is more straightfoward, as it had a uniform spacing.
With our tweaked, $200$-point grid in the linear case we could converge to $10^{-3}-10^{-4}$ accuracy, all the error coming from the trapezoidal rule in the dispersion integral. With Bernstein polynomials, we obtained a clear uniform convergence as we increased the grid size.

In the both iterations, fixed-point and Newton's method, both interpolation strategies would exhibit, near the edge of convergence, jagged solutions with stable $O(1)$ oscillations growing to similar oscillations as those displayed in figure \ref{fig:instarho}. Eventually, the oscillations become so large that the S-matrix does not satisfy unitarity anymore. While spikes could have been expected with linear interpolants, which have no notion of smoothness, this is more suprising for the Bernstein polynomials which could have been smoother. 

Finally, in terms of performance, linear interpolants performed better for the Newton method. Bernstein polynomials performed better with fixed-point iteration.

\subsection{Extension to higher dimensions}
\label{sec:extens-high-dimens}

The most exciting aspect of the method described in the present paper is that it generalizes to higher dimensions.
Indeed, an algorithm similar to the one described in section \ref{sec:atkinson-solver} can be numerically implemented in four dimensions \cite{WIP}.\footnote{Remarkably, such an attempt was made by Boguta  in 1974~\cite{Boguta:1974bm}. Unfortunately, the system of equations analyzed in that paper was not the physical one, and only two iterations were performed which overall makes the results of the analysis inconclusive.}
This refers both to the fixed-point iterations, as well as to Newton's method.

Technically, the main difference is that higher-dimensional amplitudes are functions of both energy and angle, or, equivalently, $s$ and $t$, which makes the problem two-dimensional. Upon discretization and interpolation the problem again reduces to multiplication of pre-computed matrices. 
  Increasing the resolution of the two-dimensional grid is computationally more costly, however, we believe it is feasible on a cluster, since most of the computations  can be parallelized.

The implications of elastic unitarity are much richer in higher dimensions, and the method described in this paper, to the best of our knowledge, is the only available tool to implement the correct structure of the support of the double spectral density as well as to satisfy the Mandelstam elastic unitarity equation. 
In particular, it would be very interesting to construct amplitude functions that both satisfy elastic unitarity, and maximize the coupling. The results of \cite{Paulos:2017fhb} suggest that the maximal coupling amplitudes have negligible inelasticity, therefore we expect that such amplitudes are excellent candidates to be approached using the methods described in the present paper. We hope to report on this in the near future. 

\subsection{Relation to perturbation theory}

\begin{figure}
  \centering
 \includegraphics{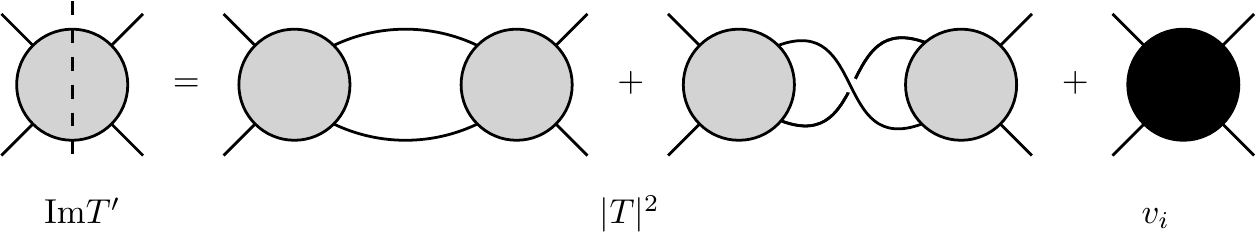}
  \caption{Graphical depiction of the fixed-point mapping action. In the RHS we included the contributions of both the $s$ and $u$-channel as required by crossing. On a given unitarity cut, $s>4$ or $s<0$, only one of them contributes. }
  \label{fig:atkison-graphical}
\end{figure}

It is instructive to compare iterations discussed in this paper to the standard perturbation theory. The diagrammatic representation of the 
iterative map is given in figure~\ref{fig:atkison-graphical}. The black dot represents inelastic processes, namely it has support only in the multi-particle region, and is kept fixed during the iteration process.  This is in sharp contrast with the usual perturbation theory where the number of new multi-particle diagrams grows factorially, see e.g. \cite{Rubakov:1995hq,Bezrukov:1995qh} and \cite{Badel:2019oxl} for a more recent discussion, and as a result the perturbation theory is typically divergent (in two dimensions, integrable theories are notable exceptions \cite{Dorey:1996gd,Gabai:2018tmm}). 

We can actually  in the iterative solution to unitarity we can relate the number we can relate the number of graphs at step $n$ to the number of graphs at step $n-1$ according to
\begin{equation}
  \label{eq:N-atkinson}
  N_{n+1} = 2(N_n)^2+1
\end{equation}
The $n$-th step iteration does not involve $n$ loop diagrams, but $L = 2^n$ loops. 
It is easy to see that \eqref{eq:N-atkinson} implies that $\log_2(N_n)\sim 2^{n} \sim L$. Therefore, we have $e^{c_0 L}$ graphs at $L$ loops which is much fewer graphs than the usual $L!$ number of Feynman graphs at $L$ loops. {In particular, denoting inelasticity by coupling $\lambda$, the series becomes $\sum_{L} \lambda^L e^{c_0 L}$ and has a finite radius of convergence given by the condition $\lambda e^{c_0}<1$.} 

While in two dimensions, the systematics of this re-summation is rather easy to understand and depict, in higher dimensions it becomes more intricate due to an additional $s-t$ crossing (as opposed to only $s-u$ crossing in two dimensions). Rather than summing one-dimensional chains of bubbles generated by iterations with beads of inelasticity $v_{i}$, higher-dimensional iterations naturally lead to two-dimensional diagrams. It would be very interesting to study them in detail. Nevertheless, we expect that the basic scaling of the number of diagrams as a function of the number of loops, with the coupling measured by inelasticity, stays the same, and therefore given that inelasticity is not too big, iterations should converge. This picture is consistent with the rigorous results of existence of nonzero convergence radius by Atkinson in four dimensions \cite{Arkinson:1968hza,Atkinson:1969wy,Atkinson:1969eh,Atkinson:1970pe}.

One immediate observation regarding the standard perturbation theory versus unitarity iterations in higher dimensions is that any fixed order in the coupling result for the scattering amplitude is not consistent with Gribov's theorem, see e.g. \cite{Correia:2020xtr} for a recent review, and thus cannot be used as an input for the convergent iterations. In other words, it is crucial for the iterative algorithm to work to have as an input a UV-improved model for inelastic effects. We leave exploration of this aspect for the future.

\subsection{Future directions}

\paragraph{Figuring out exact bounds of Atkinson's convergence in 2d}
\label{sec:figuring-out-exact}
\

In the present paper we analyzed convergence of the iterative map \eqref{eq:fixedpointiter}-\eqref{eq:schem-atk3} numerically in the discretized setting. 
It would be interesting to establish convergence criteria of the functional map directly, in the spirit of Atkinson's proofs~\cite{Arkinson:1968hza,Atkinson:1969wy,Atkinson:1969eh,Atkinson:1970pe}. We were not able to generalize the proof technique of the original papers in four dimensions to the two-dimensional case.

Atkinson's proof is nicely exemplified by the forward limit of the amplitude iteration in the lectures \cite{Atkinson:1970zz}. It proceeds in two steps. Firstly, one proves the existence of an open set of functions that maps to itself under the map which gives existence of a solution. Then one shows that the map is contracting within this set, which ensures uniqueness of this solution.

The problem we encounter to reproduce the proof in 2d is that the map involves a factor of $1/\sqrt{1-x}$, while in 4d, the square root is in the numerator $\sqrt{1-x}$.\footnote{In $d$ dimensions, the factor is $(s-4)^{(d-3)/2}\sim (1-x)^{(d-3)/2} $.} Defining $\rho'=\Phi(\rho)$, for the first step, we want to compare $||\rho'||=||\Phi(\rho)||$ to $||\rho||$, where the norm is the so-called H\"older norm, and is given by
\be
|| \rho(x) || = \sup_{0\leq x_1, x_2 \leq 1} {|\rho(x_1) - \rho(x_2) | \over |x_1 - x_2|^\mu}\,,
\ee
for some parameter $0<\mu<1$.
The factor $1/\sqrt{1-x}$ in the unitarity condition, however, seems to allow the function $\rho'$ to grow parametrically bigger than $||\rho||$ near $x=1$. Where Atkinson could find a way out by simply defining the set as the ball of radius $b$: $||\rho||<b$, we need to specify more data about the function to ensure that $||\rho||<b\implies ||\rho'||<b$.

Looking back, we know that the fixed-point iteration is divergent for 1-pole 3-zero amplitudes, albeit it resembles a lot the convergent 1-pole 1-zero amplitudes, both from the perspective of the near threshold behaviour, eq.~\eqref{eq:near-threshold} and the fast that they have similar Holder norms. Therefore, it is maybe not surprising that a generic proof \textit{\`a la} Atkinson, which assumes nothing but bounded Holder norms, should fail without adding further assumptions on the derivatives of the functions for instance.

It would be interesting to understand this in detail and rigorously establish the space of inelasticities and input parameters that lead to convergent iterations for the continuous map.

\paragraph{Theories with multiple stable particles}

An obvious direction to generalize the present analysis is to consider theories with several stable particles, e.g. theories with nontrivial flavor symmetry \cite{He:2018uxa,Cordova:2018uop,Paulos:2018fym,Cordova:2019lot}. In this case the analog of the analytic solution \eqref{eq:solution} is not readily available and the numerical techniques developed in the present work might be useful. 

\paragraph{Theories with massless particles}

In theories with massless particle there is no separation in energy between the elastic and inelastic contribution, both starting at $s=4 m^2$.
Nevertheless we can still write the unitarity equation as 
\be
{\rm Disc}_s T_{2\to2}  -  | T_{2\to2} |^2 = \text{Multi-particle} .
\ee
Then, we can again ask: given a particular form of inelasticity, can we construct an amplitude function which satisfies unitarity and crossing? It would be very
interesting to understand if the iterative techniques of the present paper can be generalized to this case. Clearly, nothing prevents us from repeating the analysis in two dimensions. In higher dimensions the situation is less clear. Such an iterative scheme would be particularly desirable for theories where inelasticity is expected to have some universality, e.g. in gravitational theories where at high energies, black holes are produced in the collision \cite{Giddings:2007qq}.

\section*{Acknowledgements}
We would like to thank Amit Sever for collaboration at
an early stage of the project. We would like to thank Miguel Correia, Andr\'{e} Martin, Slava Rychkov, and Amit Sever for useful discussions on related topics. PT is grateful to CERN for hospitality during the final phase of the project.
This project has received funding from the European Research Council (ERC) under the European Union's Horizon 2020 research and innovation programme (grant agreement number 949077).

\appendix

\section{Linear interpolation}
\label{sec:linear-interpolation}

All the plots which show iterations of both algorithms have been produced with the following grid, with 197 elements:
\begin{multline}
  \label{eq:grid-def}
  \mathtt{ Join[(3/2)^{-Range[10, 15]},Range[0, 1, 10^{-2}],Range[0, 1/10, 10^{-2}],Range[7/10, 8/10, 0.5\, 10^{-2}],}\\
    \mathtt{Range[8/10, 1, 0.25\, 10^{-2}],(1 - (3/2)^{-Range[11, 30]})] // 
    DeleteDuplicates // Rationalize // Sort
 }
\end{multline}
We provide this grid in an ancilliary file \texttt{pts196.dat}.
On this grid, we defined our linear interpolant throughout the whole domain $[0;1]$. At step $n$ we defined
  \begin{equation}
    \label{eq:rho-discrete}
    \rho_n(x) = \rho_{n,i-1} + ( \rho_n- \rho_{n,i-1})\frac{x-x_{i-1}}{x_i-x_{i-1}}\,, \quad 0\leq x_{i-1}<x<x_{i}\leq1
  \end{equation}
  where
  \begin{equation}
    \label{eq:rhoni}
    \rho_{n,i}=\rho_n(x_i)
\end{equation}
 
The next step is the dispersion integral of the interpolating function. We can compute analytically this integral on each segment $[x_i;x_{i+1}]$, in the spirit of \cite{Paulos:2016but}. Care must be taken of the position of the point $x_j$ at which is evaluated the dispersion integral, to remove logarithmic divergences at the boundaries of the segment as prescribed by the principal value integral. This lead to the definition of a matrix $B_{i,j}$ such that
  \begin{equation}
    \label{eq:B-def}
    \frac {P.V.} \pi\int_0^1 \rho_n(x)K(x,x_j)dx =\frac1 \pi \sum_{i=0}^N B_{j,i}\rho_{n,i}
  \end{equation}
where
  \begin{equation}
    \label{eq:K-def}
    K(x,x_0)=\frac2x +\frac 1{x-x_0}-\frac 1{x+\frac{x_0}{1-x_0}}
\end{equation}
is the integration kernel resulting from changing $s\to x=4/s$ in the dispersion integral. This step is the essential part of our implementation of the Atkinson program in 2d, which reduces the complexity of computing integrals at each step to a matrix action on a vector. This very same step renders the calculations \cite{Paulos:2016but} amenable to efficient numerics.

To be complete, we should also mention that we developped a mixed-interpolation strategy, using square-root interpolants, in order to have better convergence properties in the $N_{tot}$ even section, whereby, close to 1, we would resort to the following interpolants.
\begin{equation}
  \label{eq:eq:rho-discrete-sqrt}
 \rho_n(x) =-\frac{\rho_{n,i-1} \sqrt{1-x_i}-\rho_{n,i} \sqrt{1-x_{i-1}}}{\sqrt{1-x_{i-1}}-\sqrt{1-x_i}}-\sqrt{1-x}\frac{ \rho_{n,i}-\rho_{n,i-1}}{\sqrt{1-x_{i-1}}-\sqrt{1-x_i}},\quad x_{i-1}<x<x_i
\end{equation}
We did not observe a significant gain in accuracy so we did not report on related results, which are essentially identical to those obtained within the pure linear interpolant. In particular, the pure-pole CDD factor is still not a convergent fixed point of any of the algorithm.

\section{Interpolation with Bernstein polynomials}
\label{app:bernstein}

Here we describe the interpolation method with Bernstein polynomials. We use this method in one-dimensional case that is relevant for the present paper but it can be straightforwardly adapted to the higher-dimensional cases as well.

Bernstein polynomials are defined as follows
\be
\label{eq:bernstein}
b_{\nu, N}(x) = {N!  \over \nu! (N-\nu)!} x^\nu (1-x)^{N-\nu} .
\ee
Given a continuous function $f(x)$ on an interval $x \in [0,1]$, we can introduce the following interpolating function
\be
f_N(x) \equiv \sum_{\nu=0}^N f \left({\nu \over N} \right) b_{\nu,N}(x).
\ee
The famous result by Bernstein then states that as $N \to \infty$, $f_n(x)$ converges to $f(x)$ uniformly in $x$
\be
\lim_{N \to \infty} f_N(x) = f(x).
\ee

Let us consider now the following ansatz for the imaginary part of the amplitude $\rho(x)$
\be
\label{eq:bernsteinsq}
\rho(x) = \sqrt{1-x} \sum_{\nu=0}^{N} \rho\left({\nu \over N} \right) b_{\nu,N}(x) .
\ee
Imposing the decay at infinity, $\rho(0)=0$, and elastic unitarity at $x=1$, or $\lim_{x \to 1} \rho(x) = 8 \delta_{N_{tot} ,\text{even}} \ \sqrt{1-x} (1 + O((1-x)))$, we can rewrite this ansatz as follows
\be
\rho(x) = \sqrt{1-x} \Big( 8 \delta_{N_{tot} , \text{even}}  b_{N,N}(x) + \sum_{\nu=1}^{N-1}\rho_\nu  b_{\nu,N}(x) \Big) ,
\ee
where the presence of $\delta_{N_{tot} , \text{even}}$ signifies the difference of the threshold behavior of the amplitude with $N_{tot}$ even versus odd. The precise coefficient $8$ is fixed by elastic unitarity, see \eqref{eq:near-threshold}.

Alternatively, we can also use (\ref{eq:bernsteinsq}) without the $\sqrt{1-x}$ factor in front. We tried both options in implementing the iterative algorithms. While the performance of both algorithms is very similar for $N_{tot}$ odd, for $N_{tot}$ even (\ref{eq:bernsteinsq}) works much better because it correctly captures the near-threshold behavior of the amplitude.

The dispersion integral relevant for iterations takes the form
\be
\cI_{\nu,N}(x) &\equiv \int_{4}^\infty d s'  \left( {1 \over s'-s} + {1 \over s'-(4-s)} \right) \left( 1 - {4 \over s'} \right)^{1/2} b_{\nu, N} \left( {4 \over s'} \right) \nn  \\
&={\Gamma(n+1) \Gamma(n - \nu + {3 \over 2}) \over \Gamma(n+{3 \over 2}) \Gamma(n-\nu+1) }{\ _2 F_1(1,\nu ,N+{3 \over 2},{1 \over x}) +  \ _2 F_1(1,\nu,N+{3 \over 2},{x-1 \over x}) \over \nu} , ~~~ x={4 \over s}.
\ee
This integral is singular for $\nu=0$, however in the present paper we work with functions that satisfy $\rho(0)=0$ therefore the integral is always effectively finite.

Using this result we can immediately write down the matrix $B_{ij}$ that enters into the discretized unitarity relation \eqref{eq:re-discrete}
\be
B_{ij} &={\rm Re} \Big[ \cI_{j,N} \left({i \over N} \right) \Big], ~~~ i=1,..,N, ~~~ j=1,..,N-1 \ . 
\ee

\section{Newton's method : oscillations and fractals.}
\label{sec:fractal-structure}
In this short section we describe two properties of Newton's method.

\subsection{Oscillations}
\label{sec:conv-oscill}
Sometimes, the Newton method resuls in oscillations of fixed amplitude. It is not surprising, and a similar phenomenon is easily observed in 1 dimension. Let $f$ be a function such that $f(x)\sim_{x\to0}|x|^\alpha$. It is known that the root at $x=0$ can be reached by the standard Newton method only for $\alpha>1/2$. For $\alpha=1/2$, one is stuck in a cycle of $2$-point oscillations, which are therefore of $O(1)$ if the starting point is $O(1)$ away from the root. For $0<\alpha<1/2$, the Newton method overshoots the root and eventually diverges. 

\subsection{Fractals}
\label{sec:fractals}

Here we give review a simple fact about the fractal structure of the Newton algorithm.
It is a well known property of the Newton method that it gives fractal bassins of attraction (Fatou sets), separated by so-called Julia curves. Newton method applied to find the roots of the polynomial $p(z)=z^3-1$ splits the complex plane in a fractal structure known as a Julia set, which we depicted below in figure~\ref{fig:Julia}.

\begin{figure}[tb]
  \centering
  \includegraphics{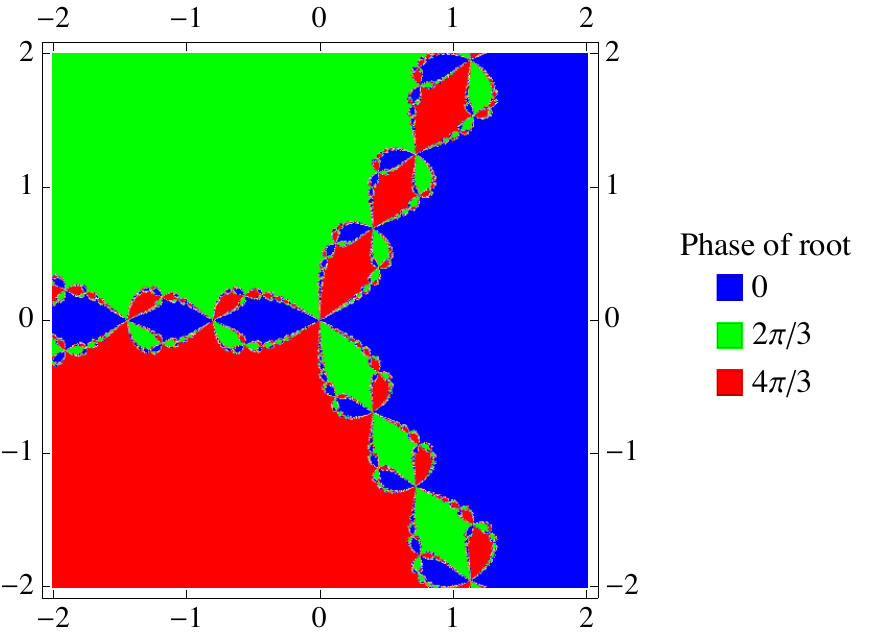}
  \caption{Plot of bassins of attraction of Newton method applied to find the roots of  $p(z)=z^3-1$. In higher dimensions, an even more complicated structure is to be expected, which is consistent with what we observe.}
  \label{fig:Julia}
\end{figure}

\section{Coupling maximization}
\label{sec:couplingmaximization}

\subsection{Coupling maximization}

It is interesting to consider the following variation of the coupling maximization problem. Consider a scattering amplitude that 
has a single bound state at the location $m_p$. Let us also assume in addition that the amplitude satisfies the following bound
at high energies
\be
\label{eq:highenergybound}
\lim_{s \to \infty} |T(s)| \leq c_0 |s|^{\alpha}, ~~~ \alpha \leq 1 ,
\ee
where $\alpha \leq 1$ is a consequence of unitarity. Following \cite{Paulos:2016but} we would like to find the amplitude that maximizes the coupling $g^2$ that is defined as the residue of the amplitude at $s=m_p^2$. 

It is easy to see that the coupling is maximized by minimizing inelasticity. Indeed, we can write \cite{Paulos:2016but}
\be
\label{eq:couplingmax}
g^2 &= g_{\text{elastic}}^2 e^{\int_{4m^2}^\infty {ds' \over 2 \pi} \log[1 - f_{i}(s')] \left( {m_p^2 (4 m^2 - m_p^2) \over s'(s'-4 m^2)} \right)^{{1 \over 2}} \left( {1 \over s' - m_p^2} + {1 \over s' - (4 m^2 - m_p^2) }  \right)} ,
\ee
where $g_{\text{elastic}}^2$ comes from $S_{\text{elastic}}(s)$ in \eqref{eq:solution}. It is clear from \eqref{eq:couplingmax} that the exponent argument is nonpositive and is minimized by setting $f_i(s)=0$. It is easy to see than that adding CDD zeros can only reduce the coupling as well, and as a result the maximum coupling is achieved by a single CDD pole $\pm S_{\text{CDD}}^{\text{pole}}(s)$.

This argument is however not valid for general $\alpha$. Indeed, a single CDD pole corresponds to (\ref{eq:highenergybound}) with $\alpha = 0$. Therefore if we want to consider amplitudes with a different behavior in the UV we should modify this amplitude. Let us argue however that such modifications do not affect the maximal value of the coupling. In other words, we can introduce a small inelasticity that can implement \eqref{eq:highenergybound} while making its contribution to \eqref{eq:couplingmax} arbitrarily small.

Let us demonstrate this very explicitly for the case $\alpha = -1$. To implement such an amplitude we would like to introduce an inelasticity that cancels the constant piece in the expansion of the CDD pole. To this extent we get the following equation
\be
\label{eq:extracondition}
\lim_{s \to \infty} T(s) &= 2\sqrt{m_p^2(4 m^2-m_p^2)} + \int_{4m^2}^\infty {ds' \over \pi} \log[1 - f_{i}(s')] \left( {1 \over s'(s'-4 m^2)} \right)^{{1 \over 2}} \left( s' - 2 m^2 \right) + O \left({1 \over s} \right) .
\ee
By taking $f_i(s) = c_0 \theta(s-s_0) {\eps \over s} ({s \over s_0})^{- \eps}$ and choosing $c_0$ appropriately, it is clear that \eqref{eq:extracondition} can be satisfied for arbitrarily small $\eps$ and arbitrarily large $s_0$. On the other hand, the contribution of such inelasticity to the coupling \eqref{eq:couplingmax} can be made arbitrarily small.

Therefore, at least in 2d we see that there are infinitely many amplitudes that satisfy all the required properties arbitrary close to the maximal value of the coupling. In some sense we can hide high energy properties of the amplitude so far in the UV such that they do not affect the IR. It would be very interesting to understand if this principle continues to hold in higher dimensions as well.

\subsection{Shift of the zero due to inelasticity}
\label{sec:shift-zero-due}

Let us consider next al situation, where the amplitude goes to a constant $\lim_{s\to \infty}T(s) = c_{\infty}$, or equivalently $S = 1+i \frac{c_{\infty}}{s} + ...$. Let us also assume that the amplitude has poles at $m_{p_i}$ and zeros at $m_{z_i}$. 

Using the general solution \eqref{eq:solution}, such asymptotic behavior leads to the following relation
\be
\label{eq:locationzero-c-inf-app}
&2\sum_{i}\sqrt{m_{p_i}^2 (4 m^2 -m_{p_i}^2)} - 2\sum_j\sqrt{m_{z_j}^2 (4 m^2 -m_{z_j}^2)} \nn \\
&= c_{\infty} - \int_{4m^2}^\infty {ds' \over \pi} \log[1 - f_{i}(s')] \left( {1 \over s'(s'-4 m^2)} \right)^{{1 \over 2}} \left( s' - 2 m^2 \right) .
\ee
The relation \eqref{eq:locationzero-c-inf} means that, if we want to keep $c_{\infty}$ and $m_{p_i}$ fixed, while turning on inelasticity, zeros has to shift towards the two-particle threshold $2m$. Indeed, since $- \log[1 - f_{i}(s')] $ increases as we increase $f_{i}(s')$ the LHS of  \eqref{eq:locationzero-c-inf} should increase as well. Given that $m_{p_i}$ are fixed the only way to achieve this is by reducing the contribution of zeros. This in turn requires moving them closer to $2m$. If the amplitude does not have any zeros, as is the case for the optimal coupling, turning on inelasticity demands to decrease $c_{\infty}$.

Let us we compare the formula above to the case of an S-matrix with one pole and zero, with the same constant at infinity but no inelasticity. For an S-matrix with a pole at $m_p$, a constant at infinity $c_{\infty}$, it is easy to see that the location of the zero $m_z'$ is given by 
\begin{equation}
  \label{eq:cinf-zero-no-inel}
 2\sqrt{m_p^2(4 m^2 -m_p^2)} - 2\sqrt{m_z'^2(4 m^2-m_z'^2)} = c_{\infty}
\end{equation}

Therefore, equating eqs.~\eqref{eq:locationzero-c-inf} and
\eqref{eq:cinf-zero-no-inel} gives that 
\begin{equation}
\label{eq:locationzero-c-infB}
2\sqrt{m_z'^2(4 m^2-m_z'^2)} - 2\sqrt{m_z^2(4 m^2-m_z^2)}=- \int_{4m^2}^\infty {ds' \over \pi} \log[1 - f_{i}(s')] \left( {1 \over s'(s'-4 m^2)} \right)^{{1 \over 2}} \left( s' - 2 m^2 \right) \geq 0 . 
\end{equation}
Adding inelasticity shifts the position of $m_z$ towards $2m$ (compared to the analogous elastic amplitude location $m_z'$),
and as a result increases the coupling.\footnote{Looking at \eqref{eq:couplingmax} one can object that increasing inelasticity has also the opposite effect of decreasing the coupling. We observe that among the two effects on the coupling shift of the zero towards $2m$ wins and overall the coupling increases as we increase inelasticity.}
The converse operation of adding inelasticity but keeping fixed the position of the zero results in
lowering the constant at infinity and the coupling in agreement with \eqref{eq:couplingmax}. These comments describe the basic behaviour of the fixed-point iteration algorithm that we observe in section \ref{sec:atkinson-numerics}.

\bibliographystyle{JHEP}
\bibliography{biblio}

\providecommand{\href}[2]{#2}\begingroup\raggedright\begin{thebibliography}{10}

\bibitem{Dorey:1996gd}
P.~Dorey, \emph{{Exact S matrices}},  in \emph{{Eotvos Summer School in
  Physics: Conformal Field Theories and Integrable Models}}, pp.~85--125, 8,
  1996, \href{https://arxiv.org/abs/hep-th/9810026}{{\ttfamily
  hep-th/9810026}}.

\bibitem{aks1965proof}
S.~O. Aks, \emph{Proof that scattering implies production in quantum field
  theory}, {\emph{Journal of Mathematical Physics} {\bfseries 6} (1965)
  516--532}.

\bibitem{Arkinson:1968hza}
D.~Atkinson, \emph{{A Proof of the Existence of Functions That Satisfy Exactly
  Both Crossing and Unitarity}: {I. Neutral Pion-Pion Scattering. No
  Subtractions.}},
  \href{http://dx.doi.org/10.1016/0550-3213(70)90120-3}{\emph{Nucl. Phys. B}
  {\bfseries 7} (1968) 375--408}.

\bibitem{Atkinson:1969wy}
D.~Atkinson, \emph{{A Proof of the Existence of Functions That Satisfy Exactly
  Both Crossing and Unitarity (Ii) Charged Pions. No Subtractions}},
  \href{http://dx.doi.org/10.1016/0550-3213(70)90121-5}{\emph{Nucl. Phys. B}
  {\bfseries 8} (1968) 377--390}.

\bibitem{Atkinson:1969eh}
D.~Atkinson, \emph{{A proof of the existence of functions that satisfy exactly
  both crossing and unitarity (iii). subtractions}},
  \href{http://dx.doi.org/10.1016/0550-3213(69)90245-4}{\emph{Nucl. Phys. B}
  {\bfseries 13} (1969) 415--436}.

\bibitem{Atkinson:1970pe}
D.~Atkinson, \emph{{A proof of the existence of functions that satisfy exactly
  both crossing and unitarity. iv. nearly constant asymptotic cross-sections}},
  \href{http://dx.doi.org/10.1016/0550-3213(70)90157-4}{\emph{Nucl. Phys. B}
  {\bfseries 23} (1970) 397--412}.

\bibitem{Atkinson:1970zz}
D.~Atkinson, \emph{{S matrix construction project: existence theorems, rigorous
  bounds and models}}, .

\bibitem{Bellazzini:2020cot}
B.~Bellazzini, J.~Elias~Mir\'o, R.~Rattazzi, M.~Riembau and F.~Riva,
  \emph{{Positive Moments for Scattering Amplitudes}},
  \href{https://arxiv.org/abs/2011.00037}{{\ttfamily 2011.00037}}.

\bibitem{Tolley:2020gtv}
A.~J. Tolley, Z.-Y. Wang and S.-Y. Zhou, \emph{{New positivity bounds from full
  crossing symmetry}},  \href{https://arxiv.org/abs/2011.02400}{{\ttfamily
  2011.02400}}.

\bibitem{Caron-Huot:2020cmc}
S.~Caron-Huot and V.~Van~Duong, \emph{{Extremal Effective Field Theories}},
  \href{https://arxiv.org/abs/2011.02957}{{\ttfamily 2011.02957}}.

\bibitem{Arkani-Hamed:2020blm}
N.~Arkani-Hamed, T.-C. Huang and Y.-t. Huang, \emph{{The EFT-Hedron}},
  \href{https://arxiv.org/abs/2012.15849}{{\ttfamily 2012.15849}}.

\bibitem{Guerrieri:2020kcs}
A.~L. Guerrieri, A.~Homrich and P.~Vieira, \emph{{Dual S-matrix bootstrap. Part
  I. 2D theory}}, \href{http://dx.doi.org/10.1007/JHEP11(2020)084}{\emph{JHEP}
  {\bfseries 11} (2020) 084},
  [\href{https://arxiv.org/abs/2008.02770}{{\ttfamily 2008.02770}}].

\bibitem{Paulos:2016fap}
M.~F. Paulos, J.~Penedones, J.~Toledo, B.~C. van Rees and P.~Vieira, \emph{{The
  S-matrix bootstrap. Part I: QFT in AdS}},
  \href{http://dx.doi.org/10.1007/JHEP11(2017)133}{\emph{JHEP} {\bfseries 11}
  (2017) 133}, [\href{https://arxiv.org/abs/1607.06109}{{\ttfamily
  1607.06109}}].

\bibitem{Paulos:2016but}
M.~F. Paulos, J.~Penedones, J.~Toledo, B.~C. van Rees and P.~Vieira, \emph{{The
  S-Matrix Bootstrap II: Two Dimensional Amplitudes}},
  \href{http://dx.doi.org/10.1007/JHEP11(2017)143}{\emph{JHEP} {\bfseries 11}
  (2017) 143}, [\href{https://arxiv.org/abs/1607.06110}{{\ttfamily
  1607.06110}}].

\bibitem{Paulos:2017fhb}
M.~F. Paulos, J.~Penedones, J.~Toledo, B.~C. van Rees and P.~Vieira, \emph{{The
  S-Matrix Bootstrap. Part Iii: Higher Dimensional Amplitudes}},
  \href{http://dx.doi.org/10.1007/JHEP12(2019)040}{\emph{JHEP} {\bfseries 12}
  (2019) 040}, [\href{https://arxiv.org/abs/1708.06765}{{\ttfamily
  1708.06765}}].

\bibitem{Homrich:2019cbt}
A.~Homrich, J.~a. Penedones, J.~Toledo, B.~C. van Rees and P.~Vieira,
  \emph{{The S-matrix Bootstrap IV: Multiple Amplitudes}},
  \href{http://dx.doi.org/10.1007/JHEP11(2019)076}{\emph{JHEP} {\bfseries 11}
  (2019) 076}, [\href{https://arxiv.org/abs/1905.06905}{{\ttfamily
  1905.06905}}].

\bibitem{Guerrieri:2020bto}
A.~Guerrieri, J.~Penedones and P.~Vieira, \emph{{S-matrix Bootstrap for
  Effective Field Theories: Massless Pions}},
  \href{https://arxiv.org/abs/2011.02802}{{\ttfamily 2011.02802}}.

\bibitem{Correia:2020xtr}
M.~Correia, A.~Sever and A.~Zhiboedov, \emph{{An Analytical Toolkit for the
  S-matrix Bootstrap}},  \href{https://arxiv.org/abs/2006.08221}{{\ttfamily
  2006.08221}}.

\bibitem{Kupsch:2008hq}
J.~Kupsch, \emph{{Towards the Saturation of the Froissart Bound}},
  \href{https://arxiv.org/abs/0801.4871}{{\ttfamily 0801.4871}}.

\bibitem{WIP}
P.~Tourkine and A.~Zhiboedov, \emph{{Scattering from production in 4d, work in
  progress}}, .

\bibitem{Symanzik:1961}
K.~Symanzik, \emph{The asymptotic condition and dispersion relations},  in
  \emph{Lectures on field theory and the many-body problem} (E.~R. Caianiello,
  ed.), ch.~10, pp.~67--92.
\newblock Academic Press, 1961.

\bibitem{Creutz:1973rw}
M.~Creutz, \emph{{Rigorous Bounds on Coupling Constants in Two-Dimensional
  Field Theories}},
  \href{http://dx.doi.org/10.1103/PhysRevD.6.2763}{\emph{Phys. Rev.} {\bfseries
  D6} (1972) 2763--2765}.

\bibitem{Mussardo:1999aj}
G.~Mussardo and P.~Simon, \emph{{Bosonic Type S Matrix, Vacuum Instability and
  Cdd Ambiguities}},
  \href{http://dx.doi.org/10.1016/S0550-3213(99)00806-8}{\emph{Nucl. Phys. B}
  {\bfseries 578} (2000) 527--551},
  [\href{https://arxiv.org/abs/hep-th/9903072}{{\ttfamily hep-th/9903072}}].

\bibitem{Vieira:TASI}
P.~Vieira, \emph{S-matrix bootstrap},  in \emph{TASI 2019 school lecture
  notes}, 2019.

\bibitem{Atkinson:1970zza}
D.~Atkinson, \emph{{Introduction to the Use of Non-Linear Techniques in
  S-Matrix Theory}},
  \href{http://dx.doi.org/10.1007/978-3-7091-5835-7_2}{\emph{Acta Phys.
  Austriaca Suppl.} {\bfseries 7} (1970) 32--70}.

\bibitem{Boguta:1974bm}
J.~Boguta, \emph{{Numerical Strategies in the Construction of Amplitudes
  Satisfying Unitarity, Analyticity and Crossing Symmetry. I}},
  \href{http://dx.doi.org/10.1016/0550-3213(74)90227-2}{\emph{Nucl. Phys. B}
  {\bfseries 72} (1974) 167--188}.

\bibitem{Rubakov:1995hq}
V.~Rubakov, \emph{{Nonperturbative aspects of multiparticle production}},  in
  \emph{{2nd Rencontres du Vietnam}: {Consisting of 2 parallel conferences:
  Astrophysics Meeting: From the Sun and Beyond / Particle Physics Meeting:
  Physics at the Frontiers of the Standard Model}}, 10, 1995,
  \href{https://arxiv.org/abs/hep-ph/9511236}{{\ttfamily hep-ph/9511236}}.

\bibitem{Bezrukov:1995qh}
F.~Bezrukov, M.~Libanov, D.~Son and S.~V. Troitsky, \emph{{Singular classical
  solutions and tree multiparticle cross-sections in scalar theories}},  in
  \emph{{10th International Workshop on High-energy Physics and Quantum Field
  Theory (NPI MSU 95)}}, pp.~228--238, 9, 1995,
  \href{https://arxiv.org/abs/hep-ph/9512342}{{\ttfamily hep-ph/9512342}}.

\bibitem{Badel:2019oxl}
G.~Badel, G.~Cuomo, A.~Monin and R.~Rattazzi, \emph{{The Epsilon Expansion
  Meets Semiclassics}},
  \href{http://dx.doi.org/10.1007/JHEP11(2019)110}{\emph{JHEP} {\bfseries 11}
  (2019) 110}, [\href{https://arxiv.org/abs/1909.01269}{{\ttfamily
  1909.01269}}].

\bibitem{Gabai:2018tmm}
B.~Gabai, D.~Maz\'a\v{c}, A.~Shieber, P.~Vieira and Y.~Zhou, \emph{{No Particle
  Production in Two Dimensions: Recursion Relations and Multi-Regge Limit}},
  \href{http://dx.doi.org/10.1007/JHEP02(2019)094}{\emph{JHEP} {\bfseries 02}
  (2019) 094}, [\href{https://arxiv.org/abs/1803.03578}{{\ttfamily
  1803.03578}}].

\bibitem{He:2018uxa}
Y.~He, A.~Irrgang and M.~Kruczenski, \emph{{A note on the S-matrix bootstrap
  for the 2d O(N) bosonic model}},
  \href{http://dx.doi.org/10.1007/JHEP11(2018)093}{\emph{JHEP} {\bfseries 11}
  (2018) 093}, [\href{https://arxiv.org/abs/1805.02812}{{\ttfamily
  1805.02812}}].

\bibitem{Cordova:2018uop}
L.~C\'ordova and P.~Vieira, \emph{{Adding flavour to the S-matrix bootstrap}},
  \href{http://dx.doi.org/10.1007/JHEP12(2018)063}{\emph{JHEP} {\bfseries 12}
  (2018) 063}, [\href{https://arxiv.org/abs/1805.11143}{{\ttfamily
  1805.11143}}].

\bibitem{Paulos:2018fym}
M.~F. Paulos and Z.~Zheng, \emph{{Bounding scattering of charged particles in
  $1+1$ dimensions}},
  \href{http://dx.doi.org/10.1007/JHEP05(2020)145}{\emph{JHEP} {\bfseries 05}
  (2020) 145}, [\href{https://arxiv.org/abs/1805.11429}{{\ttfamily
  1805.11429}}].

\bibitem{Cordova:2019lot}
L.~C\'ordova, Y.~He, M.~Kruczenski and P.~Vieira, \emph{{The O(N) S-matrix
  Monolith}}, \href{http://dx.doi.org/10.1007/JHEP04(2020)142}{\emph{JHEP}
  {\bfseries 04} (2020) 142},
  [\href{https://arxiv.org/abs/1909.06495}{{\ttfamily 1909.06495}}].

\bibitem{Giddings:2007qq}
S.~B. Giddings and M.~Srednicki, \emph{{High-energy gravitational scattering
  and black hole resonances}},
  \href{http://dx.doi.org/10.1103/PhysRevD.77.085025}{\emph{Phys. Rev. D}
  {\bfseries 77} (2008) 085025},
  [\href{https://arxiv.org/abs/0711.5012}{{\ttfamily 0711.5012}}].

\end{thebibliography}\endgroup

\end{document}